\pdfoutput=1
\documentclass[aps,prb,eqsecnum,showpacs,twocolumn]{revtex4}
\usepackage{graphicx}
\usepackage{amssymb}
\usepackage{amsmath}
\usepackage{color}
\usepackage{dsfont}
\usepackage{paralist}
\usepackage{bm}
\usepackage{verbatim}

\makeatletter
\newsavebox{\@brx}
\newcommand{\llangle}[1][]{\savebox{\@brx}{\(\m@th{#1\langle}\)}%
  \mathopen{\copy\@brx\kern-0.5\wd\@brx\usebox{\@brx}}}
\newcommand{\rrangle}[1][]{\savebox{\@brx}{\(\m@th{#1\rangle}\)}%
  \mathclose{\copy\@brx\kern-0.5\wd\@brx\usebox{\@brx}}}
\makeatother

\def\cD{{\cal D}}
\def\cH{\hat{\cal H}}
\def\hh{{\hat{h}}}
\def\cL{{\cal L}}

\def\bk{{\bf k}}

\def\bK{{\bf K}}

\def\bp{{\bf p}}
\def\br{{\bf r}}

\def\hbsigma{\hat{\boldsymbol \sigma}}

\def\holOne{\mathds{1}}

\def\hk{{\hat k}}

\renewcommand{\Im}{\mathrm{Im}}
\renewcommand{\Re}{\mathrm{Re}}

\begin{document}

\title{Unconventional localisation transition in high dimensions}

\author{
S.V.~Syzranov, V.~Gurarie, L.~Radzihovsky
} 
\affiliation{Physics Department, University of Colorado, Boulder, CO 80309, USA}

%%%%%%%%%%%%%%%%%%%%%%%%%%ABSTRACT%%%%%%%%%%%%%%%%%%%%%%%%%%%%%%%%%%%%%%%%%%%%%%%%%%%%%%%%%%%%%%%%%%%%%%%%%%%%
\begin{abstract}
  {We study non-interacting systems 
   with a power-law quasiparticle dispersion
   $\xi_{\bf k}\propto
    k^\alpha$ and a random short-range-correlated potential. 
   We show that, unlike the case of lower dimensions, for $d>2\alpha$ there exists 
   a critical disorder strength (set by the band width), at which the system exhibits a disorder-driven quantum phase transition at the bottom of the band, that lies in a universality class
   distinct from the 
   Anderson transition.
    In contrast to the conventional wisdom, it manifests itself in, e.g., the disorder-averaged 
    density of states.
	 For systems in symmetry classes that permit localisation, the striking signature 
	of this transition
	is a non-analytic behaviour
	 of the mobility edge, that is pinned to the bottom of the band for subcritical disorder and grows for disorder
	 exceeding a critical strength.     
   Focussing on the density of states, we calculate the critical behaviour (exponents and scaling functions)
   at this novel transition, using a renormalisation group, controlled by an $\varepsilon=2\alpha-d$ expansion.  
   We also
	apply our analysis to Dirac materials, e.g., Weyl semimetal, where this transition takes
   place in physically interesting three dimensions.}
\end{abstract}
%%%%%%%%%%%%%%%%%%%%%%%%%%%%%%%%%%%%%%%%%%%%%%%%%%%%%%%%%%%%%%%%%%%%%%%%%%%%%%%%%%%%%%%%%%%%%%%%%%%%%%%%%%%%%%%%


%72.10.Fk	Scattering by point defects, dislocations, surfaces, and other imperfections (including Kondo effect)
%72.15.Rn	Localization effects (Anderson or weak localization) 
  %in 72.15.-v	Electronic conduction in metals and alloys
%72.20.-i	Conductivity phenomena in semiconductors and insulators
%72.80.Vp	Electronic transport in graphene
%64.60.ae	Renormalization-group theory
  %in 64.60.A-	Specific approaches applied to studies of phase transitions
%64.60.F-	Equilibrium properties near critical points, critical exponents
  %also there
%03.65.Vf	Phases: geometric; dynamic or topological
%72.10.-d	Theory of electronic transport; scattering mechanisms
%72.20.Ee	Mobility edges; hopping transport

\date{\today}
\maketitle

\section{Introduction}

Decades of studies of transport and metal-insulator transitions in disordered materials
have resulted in well-established qualitative pictures of these phenomena\cite{AGD,Efetov:book,Kamenev:book}.
The conventional wisdom prescribes that
single-particle transport and localisation phenomena can be understood
by considering electron scattering only close to the Fermi surface;
elastic scattering through states far from the Fermi surface is believed to only finitely 
renormalise the parameters of the low-energy excitations, without any qualitative consequences.

However, such qualitative picture is not always correct, as has been known since 
the pioneering works~[\onlinecite{DotsenkoDotsenko,Fradkin1,Fradkin2}], which showed
that transport and localisation in materials with 
Dirac quasiparticle dispersion are qualitatively affected by elastic scattering between all states,
even far from the Fermi surface.
For example, in three-dimensional 
(3D) Dirac materials the scattering contribution from the full band is known to lead to a disorder-driven
phase transition between weak- and strong-disorder phases\cite{Fradkin1,Fradkin2}.
This picture has been extensively elaborated on and is now widely accepted\cite{LudwigFisher,Nersesyan:dwave,Goswami:TIRG,AleinerEfetov,Syzranov:Weyl,OstrovskyGornyMirlin,
Moon:RG,Herbut}.

In our recent paper \cite{Syzranov:Weyl} we have demonstrated
[in a controlled renormalisation-group (RG) analysis]
that such single-particle interference %phenomena
far from the Fermi surface is not specific to Dirac materials;
it dramatically affects transport and the metal-insulator transition in {\it any}
semiconductor or a semimetal in {\it sufficiently high dimensions}. This applies, in particular,
in the case of the quadratic quasiparticle dispersion
in dimensions $d\geq4$ and in the case of Dirac Hamiltonians
in dimensions $d\geq2$.

\begin{figure}[htbp]
	\centering
	\includegraphics[width=0.4\textwidth]{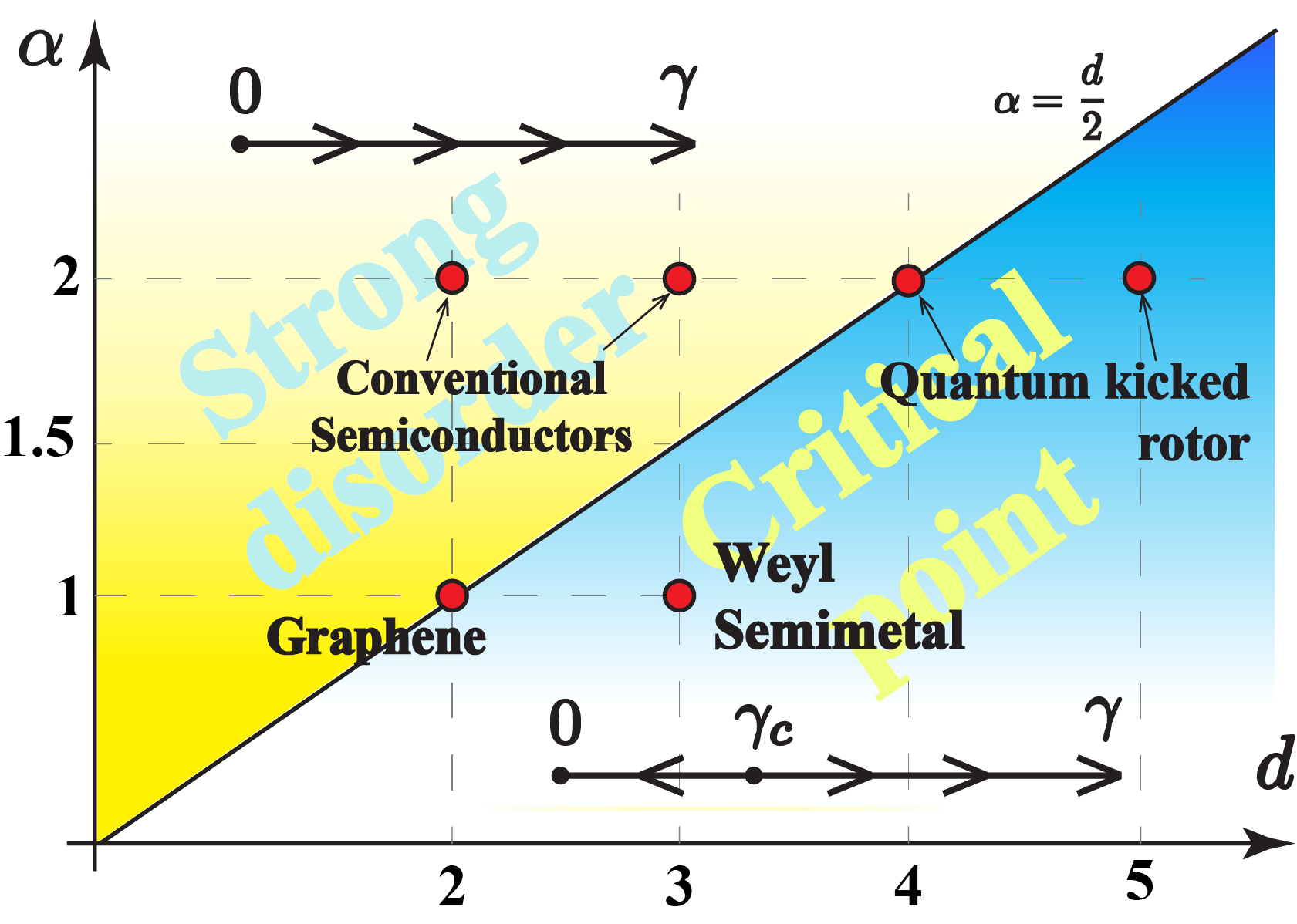}
	\caption{\label{PhaseDiagram}
	(Colour online) Critical behaviour of weak short-correlated
	disorder in materials with power-law quasiparticle
	dispersion $\xi_\bk\propto k^\alpha$ in dimension $d$. In low dimensions $d<2\alpha$ the effects
	of disorder grow at small momenta (strong-disorder regime),
	while in high dimensions $d>2\alpha$ there is a quantum phase transition between the strong-disorder
	and weak-disorder regimes. The insets show the RG flow of the dimensionless
	measure of the disorder strength relative to kinetic energy 
	$\gamma(K)\sim [K\ell(K)]^{-1}$ with decreasing the characteristic momentum $K$, where $\ell(K)$ is the mean free path.
	}
\end{figure}

As discussed in Ref.~\onlinecite{Syzranov:Weyl}, in a material with quasiparticle kinetic energy 
\begin{equation} 
\label{eq:spectrum}
\xi_\bk=ak^\alpha
\end{equation}
in the dimensions $d> d_c\equiv 2\alpha$ quasiparticle states
near the bottom of the band 
experience renormalisations from all the other states 
in the band in the presence
of a random short-range-correlated potential. 
This
leads to a quantum phase transition already in the single-particle properties as a function
of the disorder strength, 
as summarised in
Fig.~\ref{PhaseDiagram}.
Depending on whether the disorder strength $\varkappa$
is above or below a critical value $\varkappa_c$,
the effects of quenched random potential grow or decrease at small momenta, respectively.

As a result, the density of states close to
the bottom of the band exhibits a critical behaviour as a function
of disorder strength, unlike its smooth dependence on both energy
and disorder strength in the more familiar case of
$d<2\alpha$.

A well-known example of materials 
corresponding to the case of $d>2\alpha$ is the
recently realised\cite{Liu:Na3Bi,Hasan:Cd3As2,Cava:Cd3As2,Yazdani:Cd3As2,Liu:Cd3As2} Weyl semimetals,
3D materials with
Dirac-type linear quasiparticle dispersion\cite{Burkov:WeylProp,Wan:WeylProp} ($d=3$, $\alpha=1$).
While localisation in a single-valley Weyl semimetal due to potential disorder is forbidden by
symmetry\cite{Wan:WeylProp, RyuLudwig:classification}, the
weak-to-strong-disorder transition persists and manifests itself in, e.g., the 
critical behaviour of the conductivity $\sigma(\varkappa)\propto|\varkappa-\varkappa_c|^{\nu(d-2)}$,
{that has been analysed microscopically
for small but finite doping in our recent paper, Ref.~\onlinecite{Syzranov:Weyl},
and also for zero doping in Refs.~\onlinecite{Brouwer:Weyl,Fradkin1,Ominato:WeylDrude}.}
{The critical behaviour of the density of states for 3D Dirac quasiparticles has been studied
in Refs.~\onlinecite{Herbut,Fradkin1,Goswami:TIRG,Ominato:WeylDrude}}.

As we have also demonstrated there,
disordered semiconductors with the conventional quadratic quasiparticle spectrum $(\alpha=2)$
in $d>4$ dimensions 
are also characterised by a critical disorder strength.
Although one might think that such predictions are of purely academic interest,
the properties of high-dimensional semiconductors are observable experimentally:
a disordered semiconductor with the quadratic spectrum
in arbitrary dimension $d$ can be  mapped\cite{Casati:mapping,Grempel:mapping} to 
a one-dimensional periodically kicked quantum
rotor, similar to those already
realised\cite{Moor:rotorrealisation,Chabe:rotorrealisation,Delande:rotorrealisation}
in cold atomic gases to simulate Anderson localisation in 1D and 3D. Thus,
such kicked rotors present a flexible experimental platform
for observing unconventional localisation physics of high-dimensional semiconductors explored here.
Also, our results can be tested in numerical simulations of Anderson localisation
transition in high dimensions\cite{Markos:review,GarciaGarcia,Slevin,Zharekeshev:4D} close to the band edge.

In contrast,
in {\it subcritical} dimensions, $d<2\alpha$, the RG analysis shows that the effects of disorder grow at smaller momenta
and are most important close to the Fermi energy.
This is consistent with the common assumption, widely
used in the literature\cite{AGD,Efetov:book,Gantmakher:book},
that one may consider
only quasiparticles near the Fermi surface when describing transport and metal-insulator
transitions in metals and conventional semiconductors.

In this paper we further study
the weak-to-strong-disorder transition in materials with $d>2\alpha$,
such as high-dimensional semiconductors and semimetals, particularly focussing
on the disorder-averaged density of states.

We conclude the Introduction by summarising our key results and experimental predictions.
Then in Sec.~\ref{Sec:Model} we introduce the model for a semiconductor with a power-law dispersion
and short-range-correlated disorder. In Sec.~\ref{Sec:Tails} we discuss 
the tails of the density of states that emerge below the edge
of the conduction band 
due to rare fluctuations of the disorder potential (Lifshitz tails). 
In Sec.~\ref{Sec:Perturbation} we develop a perturbation theory for the states in
the conduction band and obtain divergent contributions to the effective disorder strength for
dimensions higher than critical. Sec.~\ref{Sec:RG} is devoted to the RG treatment
of the problem, controlled by an $\varepsilon=2\alpha-d$-expansion.
In Sec.~\ref{Sec:DoS} we study the disorder-averaged density of states,
the mobility threshold, and the localisation length
using scaling analysis and complementary microscopic calculations.
Sec.~\ref{Sec:Weyl}
deals with the density of states in Weyl semimetal.
We conclude
in Sec.~\ref{Sec:outlook} with a summary and a discussion of open questions.

\subsection*{Summary of the results}

The key features 
 of our findings 
for
quadratically and linearly dispersing semiconductors and Dirac semimetals %presented in this Section
is encoded in the diagram in Fig.~1. As summarised there,
for $d < 2\alpha$ the effects of random potential grow at long wavelengths
relative to the kinetic energy, which,
if allowed by symmetry and if $d>2$ (in addition to $d<2\alpha$),
leads to a mobility threshold between low-energy
localised and high-energy delocalised states.
In stark contrast, for $d > 2\alpha$ and disorder strength $\varkappa$ weaker than the critical $\varkappa_c$,
the effective disorder strength decreases relative to the kinetic energy for low momenta.
On the other hand, disorder stronger than critical grows at long wavelengths,
leading to a finite density of states and localisation (if permitted by symmetry).
This leads to a disorder-driven quantum phase transition that underlies all our results.

We study this transition 
using scaling analysis and a complementary microscopic
calculation, based on the RG-analysis, controlled by
\begin{equation}
	\varepsilon=2\alpha-d
\end{equation}
expansion, and compute a number of physical observables.

\subsubsection{Density of states}

\begin{figure}[h]
	\centering
	\includegraphics[width=0.44\textwidth]{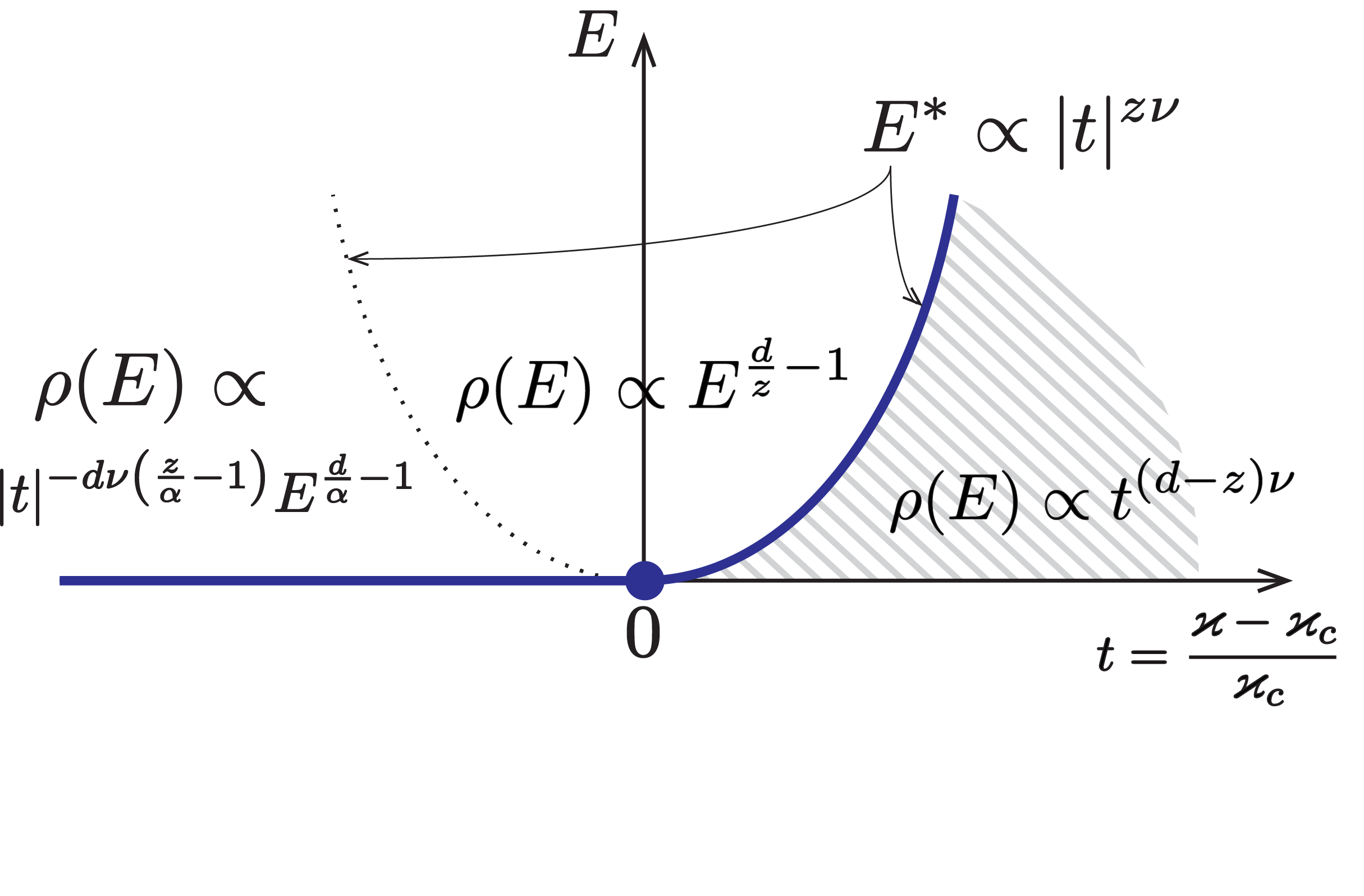}
	\caption{\label{Fig:CritDos} (Colour online)
	The energy ($E$, in the conduction band) vs. disorder strength ($\varkappa$) phase diagram for 
	a semiconductor
	in the orthogonal symmetry class (permitting Anderson localisation)
	above the critical dimension, $d>2\alpha$.
	The disorder-averaged density of states in different regimes is indicated.
	The parameter $t=\varkappa/\varkappa_c-1$ is the deviation
	of the disorder strength $\varkappa$ from the critical value $\varkappa_c$.
	The hatched region corresponds to localised states if $d>2$.
	The mobility threshold is shown as the blue curve.
	The dotted curve indicates a crossover from the critical to the effective disorder-free regime.
	}
\end{figure}

We find that the density of states exhibits the critical behaviour
{(which has been proposed previously in Ref.~\onlinecite{Herbut} for 3D Dirac quasiparticles)}
\begin{equation}
	\rho(E,\varkappa)=E^{\frac{d}{z}-1}\Phi\left[(\varkappa-\varkappa_c)/E^\frac{1}{z\nu}\right],
	\label{RhoGeneral}
\end{equation}
with the limiting cases summarised in 
Fig.~\ref{Fig:CritDos}.
Here $z$ and $\nu$ are respectively the dynamical and the correlation length critical %universal
exponents and $\Phi[x]$ is a universal scaling function.

There are three different regimes of the critical behaviour of the density of
states $\rho(E,\varkappa)$.

\begin{figure}[htbp]
	\centering
	\includegraphics[width=0.4\textwidth]{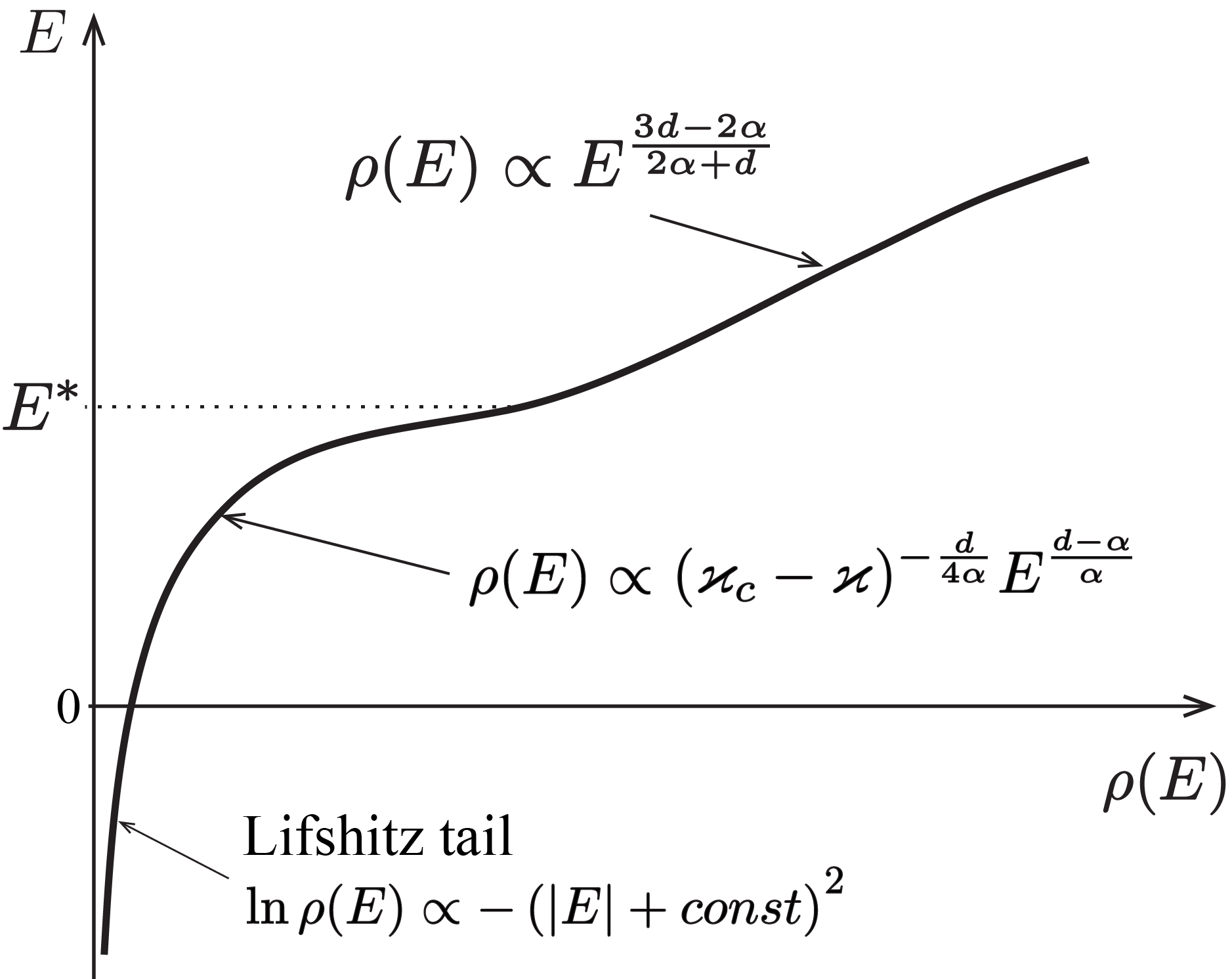}
	\caption{\label{DoSplot}
	The low-energy density of states in a disordered semiconductor in a dimension $d$ above $2\alpha$
	for subcritical disorder strength ($\varkappa<\varkappa_c$).
	}
\end{figure}

Close to the critical disorder strength, $\varkappa\approx\varkappa_c$,
the density of states is given by a power-law $\rho(E,\varkappa)\propto E^{\frac{d}{z}-1}$ and is
disorder-strength-independent. For low energies and subcritical disorder ($\varkappa<\varkappa_c$)
the energy dependence of the density of states coincides with that of a disorder-free system,
but with a disorder-dependent enhancing prefactor that diverges as the transition is approached
($\varkappa\rightarrow\varkappa_c-0$):
$\rho(E,\varkappa)\propto (\varkappa_c-\varkappa)^{-d\nu(z/\alpha-1)}E^{d/\alpha-1}$, Fig.~\ref{DoSplot}.
For low energies and strong disorder ($\varkappa>\varkappa_c$), the 
density of states is smeared by disorder and is thus finite and only weakly energy-dependent.

For a semiconductor [a material with the quasiparticle dispersion (\ref{eq:spectrum}) in the orthogonal symmetry class] 
we find the critical exponents $z$ and $\nu$ in the RG framework
with small $\varepsilon$ {in the one-loop approximation}:
\begin{align}
	\nu=-\varepsilon^{-1},
	\label{nu}
	\\
	z=\alpha-\frac{\varepsilon}{4}.
	\label{z}
\end{align}
{For instance, in the case $\alpha=2$, $d=5$, which can be 
particularly easily realised numerically, using the tight-binding model on a square lattice,
and also simulated in quantum-kicked-rotor systems,
Eqs.~(\ref{nu}) and (\ref{z}) give $\nu=1$ and $z=9/4$. We emphasise, however, that the respective
$\varepsilon=-1$ is not small and may require a similar high-loop calculation to accurately describe
experimentally and numerically observed values of the exponents $\nu$ and $z$.}

Disorder not only affects the states in the conduction band of a semiconductor, but also
leads to the formation of ``Lifshitz tails''\cite{Lifshitz:tail,ZittartzLanger,HalperinLax,LifshitzGredeskulPastur},
deeply localised states below the edge of the conduction band that occur due to rare fluctuations
of the disorder potential. 

We find that the nature of the Lifshitz tail depends crucially on whether or not the dimension $d$
is above or below critical, in the case of Gaussian disorder considered in this paper.
Unlike the conventional case of low dimensions,
broadly studied in the literature\cite{Lifshitz:tail,ZittartzLanger,HalperinLax,LifshitzGredeskulPastur},
for $d\geq2\alpha$ the density of states just
below the edge of the conduction band is exponentially suppressed at weak disorder and weakly depends on energy:
\begin{equation}
	\rho_{\text{Lifshitz}}(0)\propto\exp\left(-\frac{A}{|\varepsilon|}\frac{\varkappa_c}{\varkappa}\right),
\end{equation}
where $A$ is a constant of order unity.

We note, that the position of the band edge is shifted upon renormalisation, and we define the energy 
$E$ in the conduction band [cf. Eq.~(\ref{RhoGeneral})], as well as Lifshitz tail relative to the renormalised
edge.

Because of the exponential suppression of the tail in the limit of weak disorder
or small $\varepsilon$, here the conduction band
can be clearly distinguished from the Lifshitz tail in these limits, and
the band edge is clearly defined
%%%%%%%%%%%%%%%%%%%%%%%%%%%%%%%%%%%%%%%%%%%%%%%%%%%%%%%%%%%%%%%%%%%%%%%%%%%%%%%%%%%%%%%%%%%%%%%
\footnote{\label{EdgeNote}Throughout the paper, by the ``bottom of the band'' we mean the disorder-renormalised
edge of the band, not accounting for the rare-regions effects that lead to the formation of
Lifshitz tails (exponentially suppressed in $1/\varepsilon$).}.
%%%%%%%%%%%%%%%%%%%%%%%%%%%%%%%%%%%%%%%%%%%%%%%%%%%%%%%%%%%%%%%%%%%%%%%%%%%%%%%%%%%%%%%%%%%%%%%
This should be contrasted with the conventional case
of low dimensions, where the contribution of the Lifshitz tail can be significant
near the bottom band, and thus the band edge is not well-defined.

\subsubsection{Mobility thresholds and localisation length}

\begin{figure}[htbp]
	\centering
	\includegraphics[width=0.4\textwidth]{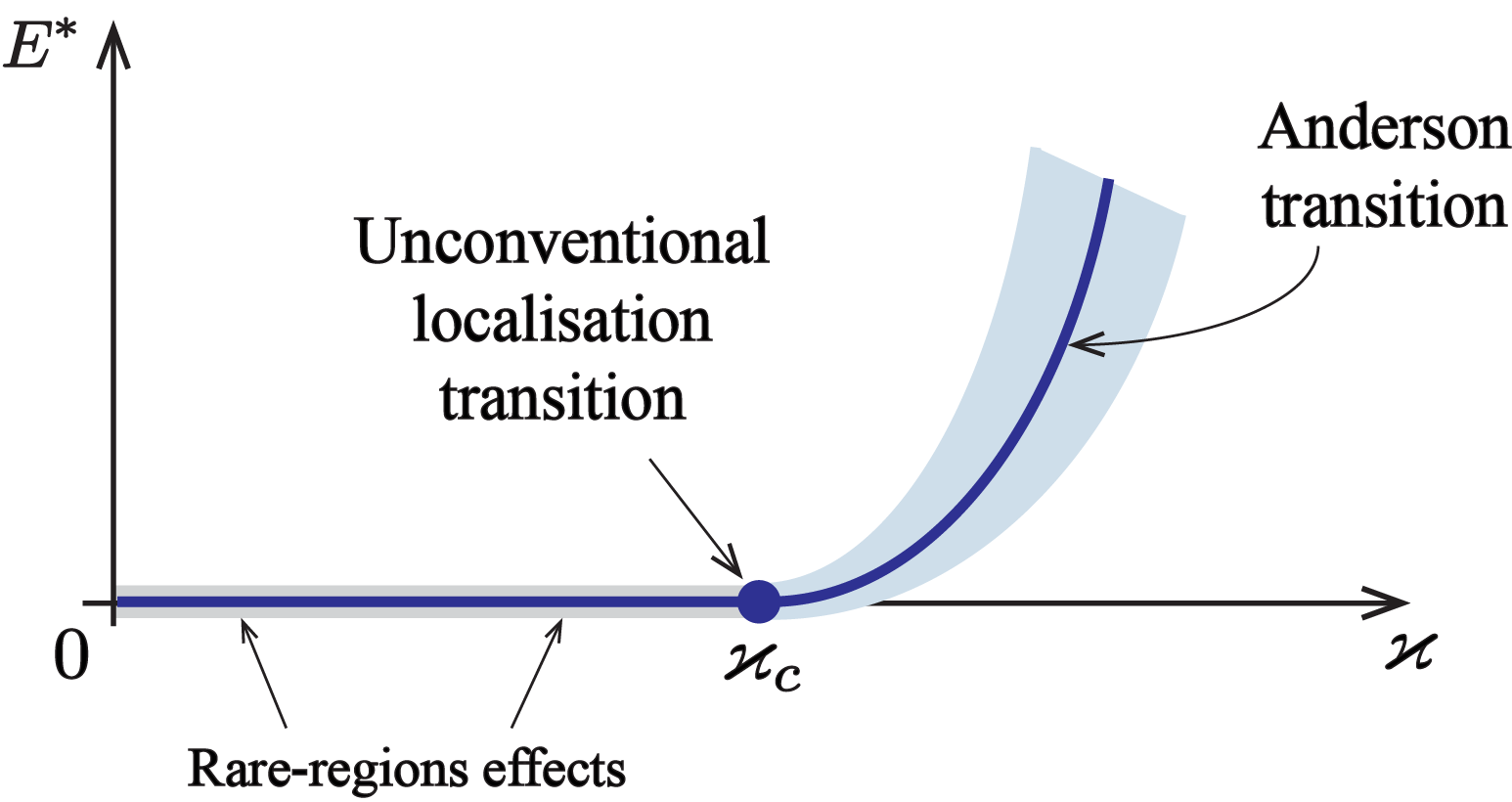}
	\caption{
	\label{Fig:MobEdge}
	(Colour online) Mobility threshold $E^*(\varkappa)$
	showing a non-analytic behaviour [given by Eq.~(\ref{MobThresh}) close to the critical point]
	as a function of the disorder strength $\varkappa$
	in higher dimensions ($d>2\alpha$). In the blue (grey) wedge-like region the critical behaviour
	is that of the Anderson universality class
	with the localisation length (\ref{XiAnderson}), and outside it is determined by
	the high-dimensional critical point ($\varkappa=\varkappa_c$, $E=0$)
	studied here.
	}
\end{figure}

Another profound consequence of single-particle interference effects in high dimensions is
the unusual behaviour of the mobility threshold [the energy $E^*(\varkappa)$
separating localised and delocalsed states]
as a function of the disorder strength, in contrast to its conventional smooth behaviour in low dimensions.
Slightly above the critical dimension ($0<-\varepsilon\ll1$) the mobility threshold is pinned to
the bottom\footnotemark[\value{footnote}]
of the band for subcritical disorder, $\varkappa<\varkappa_c$, and rapidly grows with
disorder strength for stronger disorder, $\varkappa>\varkappa_c$, as illustrated in Figs.~\ref{Fig:MobEdge}
and \ref{Fig:CritDos}.

Furthermore, we show that the critical properties of the localisation transition
in higher dimensions are richer than those below the critical dimension.
According to the conventional wisdom,
in the vicinity of the Anderson transition localised wavefunctions are characterised by a
localisation length that
diverges at the transition as
\begin{equation}
	\xi_{\text{loc}}(E,\varkappa)\propto\left|E^*(\varkappa)-E\right|^{-\nu_A}
	\label{XiAnderson}
\end{equation}
with a finite mobility threshold $E^*$ in the conduction band and
a correlation-length exponent $\nu_{A}$ which is believed to be
universal and to depend only on the dimension $d$
{and the symmetry class, provided the latter allows for localisation.}
{In particular, the exponent $\nu_A$ is believed to be independent of the
quasiparticle dispersion in a given symmetry class.}

In contrast, we find that the phenomenology in high dimensions ($d>2\alpha$) is richer.
For $\varkappa>\varkappa_c$, the critical behaviour is indeed described by
Eq.~(\ref{XiAnderson}) with a universal
exponent $\nu_A$, but with the mobility threshold vanishing as $\varkappa$ approaches $\varkappa_c+0$
[see also Eq.~(\ref{RhoGeneral})],
\begin{equation}
	E^*(\varkappa)=c(\varkappa-\varkappa_c)^{z\nu},
	\label{MobThresh}
\end{equation}
and remaining zero for $\varkappa<\varkappa_c$. 
For $\varkappa=\varkappa_c$, however,
the localisation length of the $E=0$ state diverges according to
\begin{equation}
	\xi_{\text{loc}}(\varkappa)\propto(\varkappa-\varkappa_c)^{-\nu}
\end{equation}
with the universal exponent $\nu$ given by Eq.~(\ref{nu})
in the limit of small $\varepsilon$.

Finally, for subcritical disorder, $\varkappa<\varkappa_c$,
the localisation length changes rapidly in a small energy interval,
in which the conduction band crosses over to the Lifshitz tail. Indeed, we demonstrate that in the conduction band
quasiparticle states are delocalised for $\varkappa<\varkappa_c$, because the disorder strength vanishes
upon renormalisation, while in the tail
the localisation length is of the order of the correlation length
of the potential.

\subsubsection{Weyl semimetal}

In addition to semiconductors with a scalar Hamiltonian of the kinetic energy of quasiparticles,
we study the density of states in Weyl semimetal, where electrons
are characterised by the Dirac-type dispersion $\cH(\bk)=v\hbsigma\cdot\bk$.

\begin{figure}[ht]
	\centering
	\includegraphics[width=0.44\textwidth]{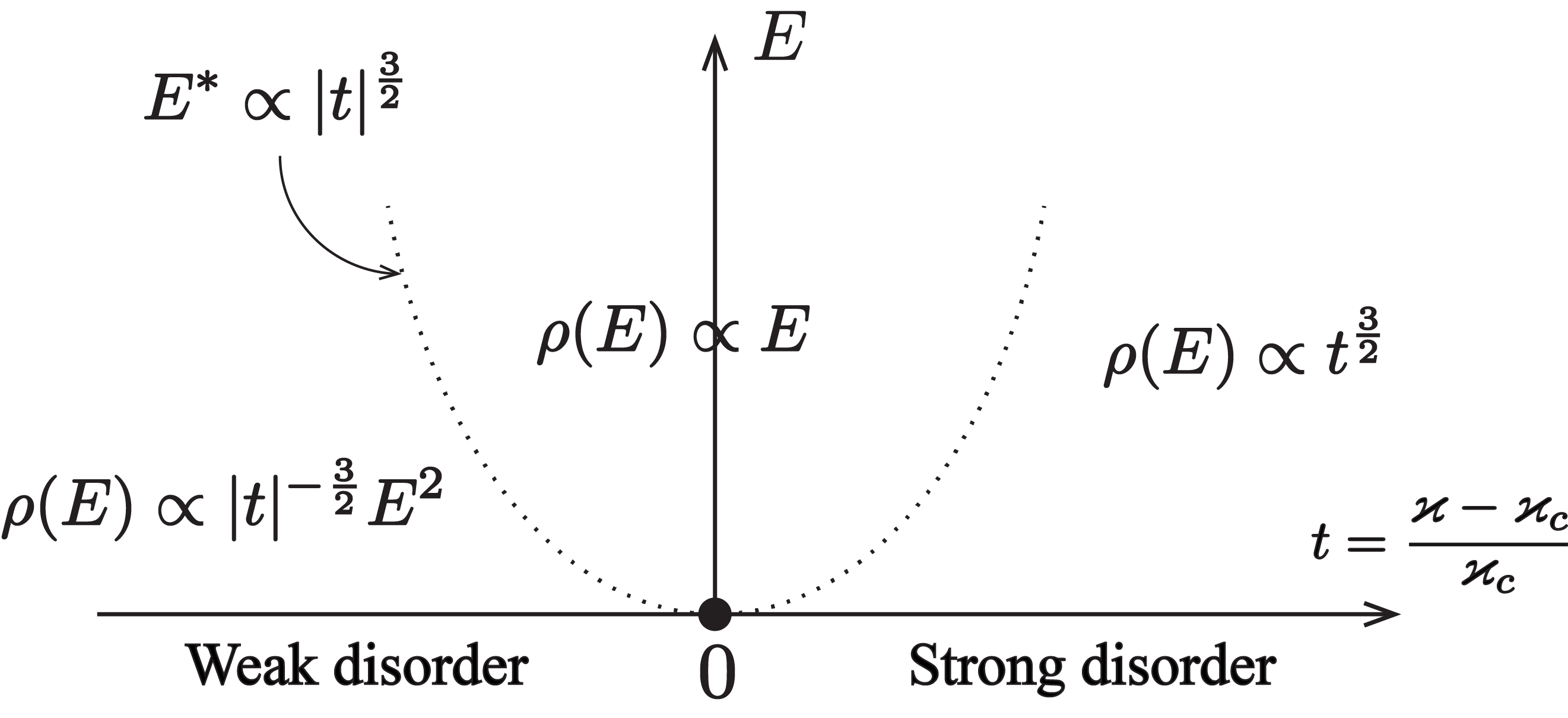}
	\caption{\label{Fig:CritDosWeyl}
	The phase diagram for disordered Weyl semimetal {($d=3$, $\alpha=1$)} illustrating weak-to-strong-disorder
	phase transition at $E=0$.	
	{Unlike a semiconductor in the orthogonal symmetry class (Fig.~\ref{Fig:CritDos}), in Weyl semimetal
	there are no localised states for sufficiently smooth disorder potential under consideration. The values
	of the exponents in the density of states $\rho(E)$ are calculated
	using a perturbative one-loop RG scheme for Dirac quasiparticles, controlled by an
	$\varepsilon=2-d$-expansion.}}
\end{figure}

\begin{figure}[ht]
	\centering
	\includegraphics[width=0.35\textwidth]{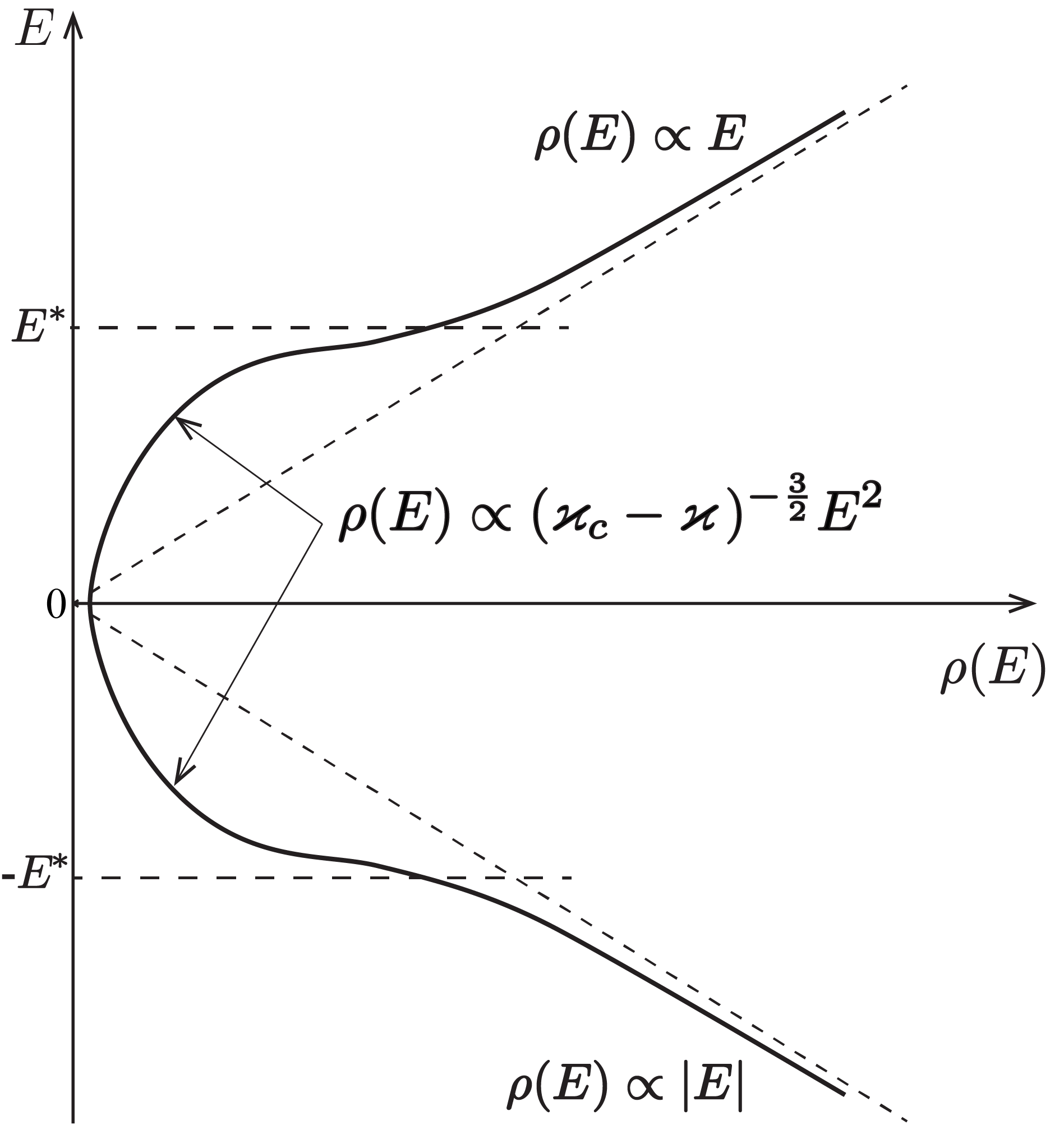}
	\caption{ \label{Fig:DoSpictureWeyl}
	The renormalised density of states in a disordered Weyl semimetal {($d=3$, $\alpha=1$)}
	near the Dirac point for subcritical disorder.
	It illustrates the crossover from the {linear}-in-E form (controlled by the 3D critical
	point close to $\varkappa=\varkappa_c$ and $E=0$) to disorder-free
	quadratic $E^2$ form at lowest energies, with
	universal 
	prefactor enhanced by disorder.}
\end{figure}

Although there is no localisation in Weyl semimetal in the presence of smooth random potential,
such system still exhibits the disorder-driven phase transition, manifested in, e.g.,
the density of states, as
summarised in Figs.~\ref{Fig:CritDosWeyl} and \ref{Fig:DoSpictureWeyl}.

These results are obtained using an RG approach (similar to the calculation for
a semiconductor) controlled by small $\varepsilon$
with $\varepsilon=-1$ set at the end of the calculation.

%%%%%%%%%%%%%%%%%%%%%%%%%%%%%%%%%%%%%%%%%%%%%%%%%%%%%%MODEL%%%%%%%%%%%%%%%%%%%%%%%%%%%%%%%%%%%%%%%%%%%%%%%%%

\section{Model}
\label{Sec:Model}

As discussed in the Introduction, in this paper we study a single-particle 
problem in the presence of a quenched random potential and analyse the effects of
the latter on the single-particle density of states and other related properties.

We consider a semiconductor with the quasiparticle Hamiltonian 
\begin{equation}
	\hh=a|\bk|^\alpha+U(\br)
	\label{Hamiltonian}
\end{equation}
in the conduction band, where $a|\bk|^\alpha$ is the kinetic energy
of a quasiparticle with momentum $\bk$, and $U(\br)$ is a weak Gaussian
disorder potential
with zero average $\langle U(\br)\rangle_{dis}=0$
and a correlation
function 
\begin{equation}
	\langle U(\br)U(\br^\prime)\rangle_{dis}
	=\Upsilon(\br-\br^\prime).
	\label{DisorderCorr}
\end{equation}
We take the latter to decay 
quickly on distances $|\br-\br^\prime|$ larger than the characteristic length $r_0$.

If the disorder potential is caused by neutral impurities, lattice defects or vacancies, $r_0$ is of
the order of the typical size of these impurities or defects.   
Disorder in semiconductors and semimetals can
be represented also by screened Coulomb impurities\cite{Shklovskii:book}, in which case $r_0$ is given by
the screening radius.
For doped semiconductors the screening is determined by the concentration of dopants,
for intrinsic semiconductors-- by electrons thermally activated from the
valence band or by electron and hole puddles that emerge due to the fluctuations of
the impurity concentration\cite{Shklovskii:book,Skinner:WeylImp,Cheianov:puddles}.

In what follows we refer to $r_0$ as the ``impurity size''. As we show in Sec.~\ref{Sec:Perturbation},
the scale $K_0=r_0^{-1}$ serves as an ultraviolet momentum cutoff for the interference effects in the conduction
band, which lead to the renormalisation of the states close to the bottom\footnotemark[\value{footnote}]
of the band.

If processes under consideration involve momentum states with wavelengths exceeding $r_0$,
the disorder
can be considered $\delta$-correlated, 
\begin{equation}
	\langle U(\br)U(\br^\prime)\rangle_{dis}=\varkappa\delta(\br-\br^\prime),
	\label{deltaDisorderCorr}	
\end{equation}
where $\varkappa=\int \Upsilon(\br)d\br$.

In this paper, in the case of a semiconductor,
we neglect electron scattering to the valence band, valid, for instance,
in the case of a sufficiently large band gap $\Delta$, separating the conduction and the valence bands,
which exceeds the width of the conduction band or the disorder-determined ultraviolet (UV)
energy cutoff $aK_0^\alpha$.

However, in the case of Weyl semimetal, there is no band gap, and we therefore take
into account scattering between the conduction and the valence bands.

Throughout the paper we assume that the dimension $d$ is integer, while the exponent $\alpha$
can be fractional. In particular, the parameter
\begin{equation}
	\varepsilon=2\alpha-d
\end{equation}
can be arbitrarily small.

%%%%%%%%%%%%%%%%%%%%%%%%%%%%%%%%%%%%%%%%%%%%%%%%%%%%%%%%%%%%%%%%%%%%%%%%%%%%%%%%%%%%%%%%%%%%%%%%%%%%%%%%%%
%%%%%%%%%%%%%%%%%%%%%%%%%%%%%%%TAILS%%%%%%%%%%%%%%%%%%%%%%%%%%%%%%%%%%%%%%%%%%%%%%%%%%%%%%%%%%%%%%%%%%%%%%
\section{Lifshitz tails and rare-regions effects in high dimensions}
\label{Sec:Tails}

While typical fluctuations of the disorder potential
can be treated in a perturbative RG analysis, discussed
in the main part of the manuscript,  
rare regions of space with large disorder potential require
more subtle non-perturbative analysis and lead to the formation of states with arbitrarily low
energies $E$ below the edge of the conduction band, known as  
``Lifshitz tail''. 
Understanding the structure
of such tails is indispensable for a complete description of single-particle states
in disordered semiconductors.

Lifshitz tails have been extensively studied for conventional
semiconductors\cite{Lifshitz:tail,ZittartzLanger,HalperinLax,LifshitzGredeskulPastur},
corresponding to low dimensions $d<2\alpha$. For quadratic quasiparticle dispersion
it has been estimated for the density of states $\rho(E)$ deep in the Lifshitz tail in dimension $d$
that
$\ln\rho(E)\propto-|E|^{2-d/2}$, in the case of Gaussian disorder, considered in this paper 
(in principle, the result is non-universal and will differ for non-Gaussian disorder).

In what immediately follows we use phenomenological arguments, similar to those of
Refs.~\onlinecite{Lifshitz:tail,ZittartzLanger,HalperinLax,LifshitzGredeskulPastur}, to obtain the density
of states
in the Lifshitz tail in high dimensions, $d>2\alpha$.
We find that the structure of the tail is dramatically different from the case of low dimensions,
again uncovering the crucial role played by the critical dimension $d_c = 2\alpha$.
In particular, for weak disorder or small $\varepsilon$ the density of states $\rho(E)$ 
weakly depends on energy, $\ln\rho(E)\approx\ln\rho(0)$, and is exponentially small for
sufficiently small energies $E$, including $E=0$, which is of particular interest to us.

The states with large negative energies $E$ occur due to rare
regions with large negative disorder potential that trap particle states.

%%%%%%%%%%%%%%%%%%%%%%%%%%More general formula for potential fluctuations%%%%%%%%%%%%%%%%%%%%%%%%%%%%%
%The distribution of the average disorder potential
%$W=\frac{1}{V}\int_\Omega U(\br) d\br$ in a spatial region $\Omega$ of volume $V$ is
%described by the Gaussian probability density
%\begin{align}
%	P_\Omega(W)= &V\left[2\pi/\iint_\Omega \Upsilon(\br-\br^\prime)d\br d\br^\prime\right]^{1/2} 
%	\nonumber\\
%	& \exp\left\{-W^2 V^2/\left[2\iint_\Omega \Upsilon(\br-\br^\prime)d\br d\br^\prime\right]\right\}
%\end{align}
%for Gaussian disorder with the correlation
%function $\Upsilon(\br-\br^\prime)$, Eq.~(\ref{DisorderCorr}).
%%%%%%%%%%%%%%%%%%%%%%%%%%%%%%%%%%%%%%%%%%%%%%%%%%%%%%%%%%%%%%%%%%%%%%%%%%%%%%%%%%%%%%%%%%%%%%%%%%%%%%

The distribution of the average disorder potential 
$W=\frac{1}{V}\int_\Omega U(\br) d\br$ in a spatial region $\Omega$ of volume $V\gg r_0^d$ is
described by the Gaussian probability density
\begin{equation}
	P_\Omega(W)=\sqrt{\frac{2\pi V}{\varkappa}}e^{-\frac{W^2V}{2\varkappa}},
	\label{deltaDistr}
\end{equation}
as follows from Eq.~(\ref{deltaDisorderCorr}).

The density of states $\rho(E)$ at a large negative energy $E$ in a semiconductor
with quasiparticle dispersion $\xi_k=ak^\alpha$
is determined by the fluctuations
of the potential on length scales $L$ that exceed the characteristic impurity size $r_0$ [the correlation
radius of the function $\Upsilon(\br)$].
We thus consider the contribution of such rare regions of deep random potential wells
of characteristic size $L$
to the density of states.

States with energy $E$ and typical linear size $L$ occur due to the potential fluctuations
$W\sim-(|E|+aL^{-\alpha})$ in spatial regions of volume $\sim L^{d}$, where $\sim aL^{-\alpha}$
estimates
the kinetic energy of the zero-point motion that raises the energy $E$ above $W$,
the bottom of the potential well.
The density of states $\rho(E)$ is determined by
the ``optimal fluctuation''~\cite{Lifshitz:tail,ZittartzLanger,HalperinLax,LifshitzGredeskulPastur}, i.e. the value
of $L$ which maximises
\begin{equation}
	\ln P_{L^d}(E)\sim -(|E|+aL^{-\alpha})^2L^d/\varkappa.
	\label{PMaximise}
\end{equation}
The dominant contribution to the density of states $\rho(E)$ is thus
determined by the competition between large scales $L$, that lower the
zero-point kinetic energy, and small scales $L$, for which potential fluctuations are more probable.

In {\it subcritical dimensions}, $\varepsilon\equiv 2\alpha-d>0$, the maximum is achieved
at $aL^{-\alpha}=|E|d/\varepsilon$, leading to the conventional result
\begin{equation}
	\rho_{\text{Lifshitz}}(E)\propto\exp
	\left[
	-C|E|^{2-\frac{d}{\alpha}}\left(1+\frac{d}{\varepsilon}\right)^2
	\left(\frac{a\varepsilon}{d}\right)^\frac{d}{\alpha}\varkappa^{-1}
	\right],
	\label{LowDTail}
\end{equation}
where $C$ is a constant of order unity. The density of states (\ref{LowDTail}) has been obtained previously
for conventional semiconductors~\cite{Lifshitz:tail,ZittartzLanger,HalperinLax,LifshitzGredeskulPastur}.

In {\it high dimensions}, $d>2\alpha$, the expression (\ref{PMaximise}) has no maximum at finite $L$
and grows infinitely as $L\rightarrow 0$. Thus, the density of states in high dimensions is determined
by the shortest microscopic length scales. The minimal scale in the model is the ``impurity size'' $r_0$.
Inserting $L\sim r_0$ in Eq.~(\ref{PMaximise}), we obtain
\begin{equation}
	\rho_{\text{Lifshitz}}(E)\propto\exp\left[
	-C_1\left(|E|+C_2ar_0^{-\alpha}\right)^2r_0^d/\varkappa
	\right].
	\label{HighDTail}
\end{equation}

Eqs.~(\ref{LowDTail}) and (\ref{HighDTail}) correctly describe the densities of states in low ($d<2\alpha$)
and high ($d>2\alpha$) dimensions respectively, provided the respective exponentials
are significantly smaller than unity.
While Eq.~(\ref{LowDTail}) thus applies in low dimensions only for sufficiently large negative
energies $|E|\gg a^{-\frac{d}{\varepsilon}}\varkappa^\frac{\alpha}{\varepsilon}$,
Eq.~(\ref{HighDTail}) describes the density of states in the Lifshitz
tail in high dimensions for all negative energies
provided the disorder is weak enough, $\varkappa\ll a^2r_0^{-\varepsilon}$.
%%%%%%%%%%%%%%%%%%%%%%%IMPORTANT%%%%%%%%%%%%%%%%%%%%%%%%%%%%%%%%%%%%%%%%%%%%%%%%%%%%%%%%%%%%%%%%%%%%%%%%%%
% The latter condition is fulfilled for
%any subcritical disorder $\gamma<\gamma_c\sim-\varepsilon a^2\ll a^2$ for small negative $\varepsilon$.
%%%%%%%%%%%%%%%%%%%%%%%%%%%%%%%%%%%%%%%%%%%%%%%%%%%%%%%%%%%%%%%%%%%%%%%%%%%%%%%%%%%%%%%%%%%%%%%%%%%%%%%%%%

Our result,
Eq.~(\ref{HighDTail}), thus
shows that in high dimensions the density of states weakly depends on energy for
$|E|\lesssim ar_0^{-\alpha}$ and decays exponentially $\rho(E)\propto\exp(-C_1|E|r_0^d/\varkappa)$ otherwise.

{\it Gapless semiconductors.} Since Eq.~(\ref{HighDTail}) applies for all energies 
below the bottom\footnotemark[\value{footnote}] of the band,
it can be used to describe the smearing of the density of states at the degeneracy point in gapless semiconductors,
i.e. materials where the conduction and the valence bands touch, such as Weyl semimetals or graphene.
In these materials there is no band gap, so the expression (\ref{HighDTail}) can be used only for $E=0$, i.e.
in the bottom of the conduction band (the top of the valence band). Indeed, for $E=0$, $d=3$, and $\alpha=1$,
Eq.~(\ref{HighDTail}) gives the density of states $\rho(0)$ in Weyl semimetal, recently
obtained in Ref.~\onlinecite{Nandkishore:rare}.

\section{Perturbation theory in the conduction band}
\label{Sec:Perturbation}

In Sec.~\ref{Sec:Tails} we addressed the density of states below the edge of the conduction
band due to {\it rare} fluctuations of the disorder potential. Let us now consider the states
in the conduction band, where it is sufficient to consider the {\it typical} fluctuations
of the random potential.

%%%%%%%%%%%%%%%%%%%%%%%%%%%%%%%%%%%%%%%%%%%%%%%%%%%%%%%%%%%%%%%%%%%%%%%%%%%%%%%%5
\subsection{Phenomenological argument for the existence of the
critical dimension $d_c=2\alpha$}
\label{Subsec:phenom}

Before turning to more technical perturbative and RG analyses, we assess 
of the role of quenched disorder using phenomenological scaling arguments.
The importance of the random potential to a singe-particle state of momentum $k$
can be assessed by comparing the kinetic energy $ak^\alpha$ with the typical 
fluctuation $U_{\text{rms}} \sim \left[\varkappa k^d\right]^\frac{1}{2}$
of the (zero-mean) random potential, averaged
over the volume $k^{-d}$, set by the de Broglie wavelength $1/k$.

For $d>2\alpha$, the ratio $U_{\text{rms}}/(ak^\alpha)=k^{d/2-\alpha}\varkappa^{1/2}/a$
[$\sim\sqrt{\gamma}$, with $\gamma$ being the dimensionless measure of disorder strength,
introduced in Eq.~(\ref{GammaDefinition}) below]
vanishes with reduced momentum, which reflects the irrelevance of disorder (in RG parlance)
in higher dimensions. 

In contrast, for $d < 2\alpha$ the effects of disorder grow at small momenta.
For $d>2$ (in addition to $d<2\alpha$), we expect the localisation
of
particles with sufficiently low momenta $k\lesssim K^*$, such that
the kinetic energy $a\left(K^*\right)^\alpha$ is of the order of
the characteristic disorder potential fluctuation $U_{\text{rms}}\left(K^*\right)$.
This allows us to estimate the mobility threshold in lower dimensions:
\begin{equation}
	E_{\text{mob}}\sim a^{1-2\alpha/\varepsilon}\varkappa^{\alpha/\varepsilon}.
	\label{MobNaive}
\end{equation}

Although the above phenomenological argument allows one to predict the existence 
of the critical dimension $d_c=2\alpha$ and qualitatively different
effects of disorder in dimensions $d>2\alpha$ and $d<2\alpha$, such argument neglects
elastic scattering of the states with characteristic momentum $k$ through
the states $K\gg k$.

We show below that such large-momentum scattering is important in higher dimensions, but
may be neglected in the dimensions below critical. Indeed, sufficiently below the critical
dimension [in the limit $\varepsilon\gg\varkappa/(a^2K_0^\varepsilon)$], the phenomenological
estimate (\ref{MobNaive}) of the mobility threshold
is accurate and coincides with the result [Eq.~(\ref{SubMobExact}) below] of a rigorous RG calculation. 
However, when approaching the critical dimension ($\varepsilon\rightarrow0$), the above
estimate, Eq.~(\ref{MobNaive}), is no longer accurate,
as elastic scattering between all states in the band needs to be taken into account.

\subsection{Perturbative correction to the disorder strength}

In what immediately follows we apply perturbation theory to show that sufficiently
below critical dimensions, $2\alpha-d\gtrsim 1$, 
the quasiparticle transport in a weakly disordered semiconductor is dominated by the scattering between
states in a narrow momentum 
shell, $|k-K|\ll k$,
whereas
close to or above the critical
dimensions, $d>2\alpha$, scattering in a large band of momenta $k < K_0$, up to the UV cutoff $K_0$,
is important.

\begin{figure}[htbp]
	\centering
	\includegraphics[width=0.25\textwidth]{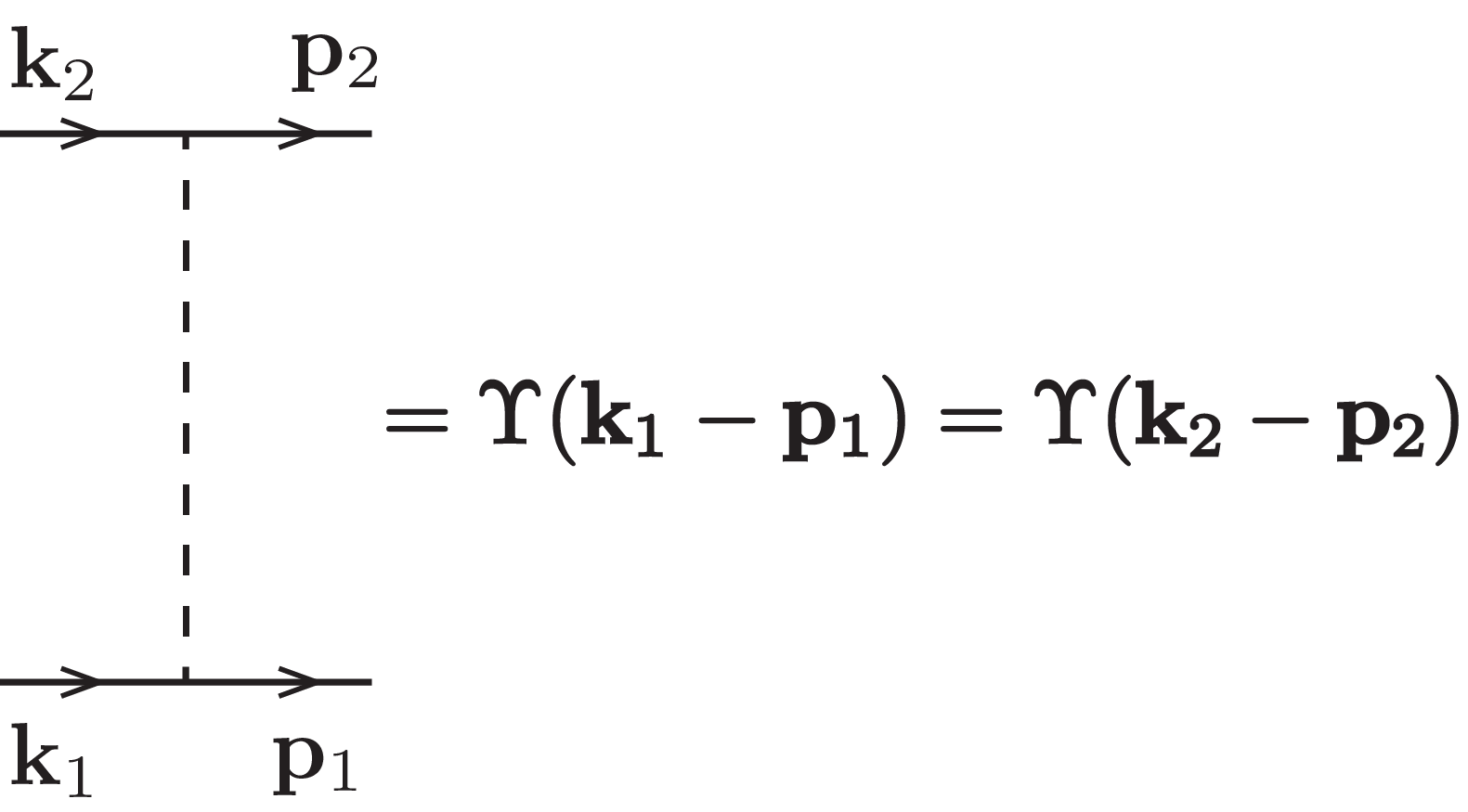}
	\caption{\label{Fig:Impline}
	Impurity line.
	}
\end{figure}

\begin{figure}[h]
	\centering
	\includegraphics[width=0.45\textwidth]{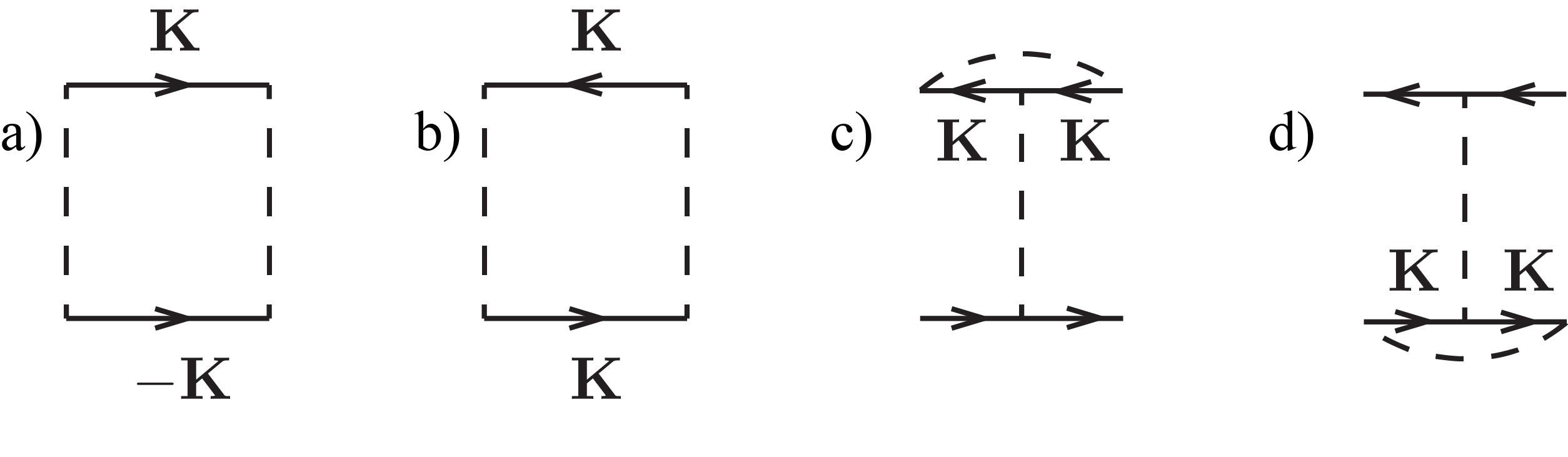}
	\caption{
	\label{Fig:LineDiagrams}
	The leading-order diagrams for the renormalisation of the impurity line due to scattering through
	states with large momenta. Large momentum $\bK$ significantly exceeds the other incoming and outgoing momenta.
	Diagrams a)-d) have equal values.
	}
\end{figure}

In the leading order in the disorder strength,
the effect of large-momentum scattering ($|k-K|\gtrsim k$) %(outside of the shell $|k-K|\lesssim l^{-1}$)
on states with small momenta $k$ 
can be illustrated by the renormalisation of the impurity line, Fig.~\ref{Fig:Impline},
mimicked diagramatically in Fig.~\ref{Fig:LineDiagrams} and estimated as
\begin{equation}
	\delta\Upsilon(\sim k)\sim 4\int_{K\gtrsim k} \frac{\Upsilon^2(\bK)}{\xi_\bK^2}\frac{d^d\bK}{(2\pi)^d},
\end{equation}
where $\Upsilon(\bK)$ is the Fourier-transform of the disorder correlation function $\Upsilon(\br)$,
Eq.~(\ref{deltaDisorderCorr}),
and $\xi_\bK=aK^\alpha$ is the kinetic energy of a quasiparticle with momentum $\bK$.

For
short-range-correlated disorder, which we consider in this paper, the function $\Upsilon(\bK)$ decays fast
beyond the cutoff momentum $K_0=r_0^{-1}$, and the renormalisation of the impurity line can be rewritten in
terms of modification of the disorder strength $\varkappa$, Eq.~(\ref{deltaDisorderCorr}),
\begin{equation}
	\delta\varkappa\sim 4C_d\frac{\varkappa^2}{a^2}\int_{k}^{K_0}\frac{dK}{K^{2\alpha-d+1}},
	\label{kappaEstimate}
\end{equation}
where $C_d=S_d/(2\pi)^d$ and $S_d$ is the area of a unit
sphere in a $d$-dimensional space.

\subsubsection{
Subcritical dimensions
}

Consistent with the phenomenological analysis of Subsection~\ref{Subsec:phenom},
in the dimensions $d<2\alpha$ the integral in Eq.~(\ref{kappaEstimate})
is dominated by momenta $K\sim k$ near the lower limit, and
\begin{equation}
	\delta\varkappa\sim \frac{1}{2\alpha-d}\frac{\varkappa}{k\ell(k)},
	\label{kappaModAnswer}
\end{equation}
where we have introduced the quasiparticle mean free path (cf. Appendix~\ref{Sec:MFP} for a detailed
calculation of the mean free path)
\begin{equation}
	\ell(k)=\frac{\alpha^2a^2 k^{2\alpha-d-1}}{2\pi C_d\varkappa}.
	\label{MeanFreePath}
\end{equation}

The quantity $k\ell(k)$, entering Eq.~(\ref{kappaModAnswer}), is an important parameter in the 
conventional Anderson localisation theory
in the dimensions $d$ (sufficiently) above $2$.
If this parameter is large, $k\ell(k)\gg1$, the respective states are delocalised, according
to the so-called Ioffe-Regel criterion\cite{Ioffe:IoffeRegel,IoffeRegel}. Otherwise, $k\ell(k)\sim 1$, and
the respective states are either localised and do not contribute to transport or are close to localisation.
In $d\leq2$ all states are localised.

Conventional semiconductors in 2D and in 3D are characterised by quadratic quasiparticle
spectrum ($\alpha=2$), Fig.~\ref{PhaseDiagram}, and thus correspond to the dimensions below critical, 
$2\alpha-d\gtrsim 1$. Then Eq.~(\ref{kappaModAnswer}) shows that for
states with $k\ell(k)\gg1$
the large-momentum scattering produces only small corrections to the disorder strength 
$\delta\varkappa\sim\varkappa/[k\ell(k)]$ and can be neglected.

Thus, in conventional semiconductors one can apply the usual transport
theory and disorder-averaging techniques\cite{AGD}, with quasiparticle
scattering confined inside a small momentum shell near the Fermi surface
and neglecting the other states in the band.

\subsubsection{
Dimensions close to or above critical
}
When approaching the critical dimension, $d\rightarrow 2\alpha$, the renormalisation of the
disorder strength~(\ref{kappaModAnswer}) by interference processes involving large momenta
dramatically increases.

In the dimensions $d>2\alpha$ the integral in Eq.~(\ref{kappaModAnswer}) is dominated by
large momenta close to the ultraviolet cutoff $K_0=r_0^{-1}$;
\begin{equation}
	\delta\varkappa=4C_d\frac{\varkappa^2}{a^2}\frac{1}{d-2\alpha}\frac{1}{r_0^{d-2\alpha}}.
\end{equation}
The modification of the disorder strength by processes involving momenta $\sim K_0$ can become very large
and diverges in the limit of $\delta$-correlated disorder $r_0\rightarrow0$.

Such effects cannot be treated perturbatively and require adequate renormalisation-group analysis,
to which we turn in the next section.

%%%%%%%%%%%%%%%%%%%%%%%%%%%%%%%%%%%%%%%%%%%%%%%%%%%%%%%%%%%%%%%%%%%%%%%%%%%%%%%%%%%%%%%%%%%%%%5
\section{Renormalisation group analysis}
\label{Sec:RG}

In order to address the effects of random potential beyond the above phenomenological
and perturbative approaches,
in this Section we develop a logarithmic renormalisation-group description for the states in the conduction
band in the critical
dimension $d=2\alpha$ and address the other dimensions by means of an 
\begin{equation}
	\varepsilon=2\alpha-d
\end{equation}
-expansion.

Similar renormalisation-group descriptions have been developed for systems
with Dirac-type quasiparticle dispersion
in two and three dimensions (2D and 3D), such as the
Ising model\cite{DotsenkoDotsenko},
integer-Hall-effect systems\cite{LudwigFisher}, d-wave superconductors\cite{Nersesyan:dwave},
topological insulators~\cite{Goswami:TIRG},
graphene~\cite{AleinerEfetov}, and Weyl semimetals\cite{Syzranov:Weyl,Fradkin1,Fradkin2}.

For concreteness and because of its central role in characterising the system,
we focus on the disorder-averaged single-particle density of states
\begin{equation}
	\rho(E)=-\frac{1}{\pi}\Im\left<\frac{1}{V}\int d\br\: G^R(\br,\br,E)\right>_{dis}
	\label{rhoGen}
\end{equation}
in the conduction band, where $\langle G^R(\br,\br^\prime,E)\rangle_{dis}$
is the disorder-averaged retarded Green's function. In the
supersymmetric representation\cite{Efetov:book} (here used as a convenient tool,
although Keldysh and replica representations can be equivalently utilised)
\begin{align}
	& \left< G^R(\br,\br,E)\right>_{dis}=-i\int\cD\psi\cD\psi^\dagger
	  e^{-\cL_0-\cL_{int}}s(\br)s^*(\br),
	  \label{Action10}
	  \\
	& \cL_0=-i\int \psi^\dagger
	\left[E+i0-a|\hk|^\alpha\right]\psi\: d\br,
	\label{Action20}
	\\
	& \cL_{int}=\frac{1}{2}\varkappa\int \left(\psi^\dagger\psi\right)^2d\br,
	\label{Action30}
\end{align}
where $\psi=(\chi,\: s)^T$ and $\psi^\dagger=(\chi^*,\: s^*)$ are a row and a column of anticommuting (Grassman)
$\chi$, $\chi^*$ and commuting $s$, $s^*$ fields, and $\hk=-i\partial_\br$. 

In Eq.~(\ref{Action30}) we have taken the random potential
to be zero-mean and $\delta$-correlated,
as the low-energy states under consideration are smooth on the scale $r_0=K_0^{-1}$.

Integrating out the modes with the highest momenta in an infinitesimal shell $K e^{-l}<k<K$
leads to a modified expression
for the density of states
\begin{align}
	\rho(E)=\frac{1}{\pi V}\Re\left[\lambda(K)
	\int\cD\psi\cD\psi^\dagger d\br\:
	  e^{-\tilde\cL_0-\tilde\cL_{int}}s(\br)s^*(\br)\right]
	\label{Action1}
\end{align}
with a renormalised Lagrangian $\tilde\cL_0+\tilde\cL_{int}$,
\begin{align}
	& \tilde\cL_0=-i\int \psi^\dagger
	\left[\lambda(K)(E+i0)-a|\hk|^\alpha\right]\psi\: d\br,
	\label{Action2}
	\\
	& 
	\tilde\cL_{int}=\frac{1}{2}\tilde\varkappa(K)\int \left(\psi^\dagger\psi\right)^2d\br,
	\label{Action3}
\end{align}
where the resulting effective couplings $\lambda(K)$ and $\tilde\varkappa(K)$ flow as
\begin{align}
	\partial_l\lambda &=\frac{C_d}{a^2}\tilde\varkappa\lambda K^{-\varepsilon},
	\label{lambdaflow0}
	\\
	\partial_l\tilde\varkappa &=\frac{4 C_d}{a^2} \tilde\varkappa^2 K^{-\varepsilon},
	\label{gammaflow0}
\end{align}
with the initial values $\tilde\varkappa(K_0)=\varkappa$ and $\lambda(K_0)=1$
(for a detailed derivation of the RG equations see Appendix~\ref{Sec:RGDetail}).

The renormalised Lagrangian retains the $\delta$-correlated disorder form,
\begin{equation}
	\langle U(\br) U(\br^\prime)\rangle=\tilde\varkappa(K)\delta(\br-\br^\prime),
\end{equation}
with $\tilde\varkappa(K)$ characterising the renormalised disorder strength.
The parameter $\lambda(K)$ plays the role of the inverse quasiparticle weight.

We note, that the edge of the conduction band also flows under the RG.
Thus, throughout the paper, the energy $E$ is implicitly understood to be measured from the 
renormalised band edge.

The form of the flow equations (\ref{lambdaflow0})-(\ref{gammaflow0})
for dimensionful couplings suggests an introduction of a dimensionless
measure of disorder strength
\begin{equation}
	\gamma(K)=\frac{4C_d}{a^2}\tilde\varkappa(K)K^{-\varepsilon},
	\label{GammaDefinition}
\end{equation}
in terms of which the RG equations reduce to a simple form
\begin{align}
	\partial_l\lambda &=\gamma\lambda/4,
	\label{lambdaflow}
	\\
	\partial_l\gamma &=\varepsilon\gamma+\gamma^2.
	\label{gammaflow}
\end{align}

The RG Eqs.~(\ref{lambdaflow}) and (\ref{gammaflow}) are similar to those for systems
with Dirac-type quasiparticle dispersion, extensively studied in the
literature\cite{DotsenkoDotsenko,Fradkin1,Fradkin2,LudwigFisher,Nersesyan:dwave,Goswami:TIRG,AleinerEfetov,Syzranov:Weyl,Moon:RG,RoyDasSarma}.
We discuss the RG equations for such Dirac materials and the critical behaviour, that follows from them,
in Sec.~\ref{Sec:Weyl}.

We note that the dimensionless parameter $\gamma(k)$ is related
to the mean free path $\ell(k)$, Eq.~(\ref{MeanFreePath}) as
\begin{equation}
	\gamma(k)=\frac{2\alpha^2}{\pi}\frac{1}{k\ell(k)}
	\label{GammaKL}
\end{equation}
and is also
a square of the ratio $U_{\text{rms}}(k)/(ak^\alpha)$ of the rms value of the random potential
to the kinetic energy at momentum $k$ (see Subsection~\ref{Subsec:phenom}).
In realistic system $\alpha\sim1$, so $\gamma(k)$ is of the order
of the parameter\cite{Ioffe:IoffeRegel,IoffeRegel}
$[k\ell(k)]^{-1}$, which plays an important role\cite{AGD,Efetov:book,Gantmakher:book} in the studies
of disordered metals and semiconductors.

Thus, the parameter $\gamma(K)$ reflects the localisation properties of the states
with momenta of the order of
 $K$ in $d>2$ dimensions (cf. also Appendix \ref{Sec:MFP}). 
Namely, according
to the Ioffe-Regel criterion (and as supported by detailed microscopic
calculations\cite{Efetov:book,Kamenev:book}), the state
with energy $E$ is delocalised if $\gamma(K_E)\ll 1$ with $K_E$ given by
Eq.~(\ref{MomTerminate}).
If disorder grows upon renormalisation, the mobility threshold is reached at the
value of the momentum cutoff $K$, such that $\gamma(K)\sim1$. 

{\it Termination of the RG.} 
To utilise our RG approach
for a computation of a physical quantity at energy $E$
[e.g., the density of states $\rho(E)$], we stop integrating out high-momentum
modes when the momentum cutoff $K$ reaches an $E$-dependent value $K_E$, such that
\begin{equation}
	\lambda(K_E) E\sim aK_E^\alpha,
	\label{MomTerminate}
\end{equation}
as determined by the quadratic part of the Lagrangian, Eq.~(\ref{Action2}).  

On the other hand our RG approach is perturbative in the dimensionless disorder strength $\gamma$
and is thus only valid for $\gamma \ll 1$.
This therefore places a condition ($E>E^*$)
on the minimum energy that can be studied within this
analysis in a  regime where disorder is relevant at low energies.

The RG procedure must
also be terminated if the density of states $\rho(E)$, derived from Eqs.~(\ref{Action1})-(\ref{Action3}),
 becomes smaller than the density of states $\rho_{\text{Lifshitz}}(0)$
in the Lifshitz tail near the edge of the band,
emerging due to rare strong fluctuations of the disorder potential. Indeed, the latter occur as
instantons in the disorder-averaged quasiparticle action\cite{Cardy:tails,Yaida:tails} and thus
cannot be taken into account by a perturbative
RG procedure. If the instanton contribution to the density of states dominates, the
contributions from the typical disorder fluctuations
are no longer important.

In high dimensions $d>2\alpha$ the density of states (\ref{HighDTail}) in the Lifshitz tail does not
experience renormalisations from the interference effects in the conduction band, because it originates
from rare fluctuations of the random potential
on the scale of the disorder correlation length $r_0$, i.e. from the momentum
modes close to the ultraviolet cutoff $K_0$.
However, the density of states (\ref{LowDTail})
just below the critical dimensions, $0<2\alpha-d\ll1$, is subject to renormalisations.

{\it Critical point.} Below critical dimensions ($\varepsilon>0$), $\gamma(l)$ always flows to larger values,
according to Eq.~(\ref{gammaflow}). 
This encodes the conventional wisdom that the effective random potential
becomes stronger at the bottom\footnotemark[\value{footnote}] of the band. For  $2 < d < 2\alpha$, this
is consistent with the usual expectation of the existence of a mobility edge
that evolves smoothly in the conduction band as a function of disorder strength.

In qualitative contrast to this conventional expectation, for supercritical dimension, $d > 2\alpha$, 
$\gamma(l)$ is irrelevant, flowing to the $\gamma = 0$ disorder-free Gaussian fixed point 
for $\gamma$ smaller than the critical value
\begin{equation}
	\gamma_c = -\varepsilon,
	\label{gammac}
\end{equation}
in accordance with the phenomenological analysis of Subsection~\ref{Subsec:phenom}.
Instead, for disorder strength exceeding the critical $\gamma_c$, $\gamma(l)$ flows to larger values,
reflecting the relevance of strong disorder in higher dimensions.
These two regimes are then separated by a critical fixed point $\gamma_c$.

Thus, for $d>2\alpha$ ($\varepsilon<0$) the renormalisation flow leads
to a disorder-driven quantum phase transition.
Namely,
the effects of the random potential on the states near the edge of the band
may be significant or negligible depending on whether or not the disorder strength $\varkappa$
exceeds the critical value
\begin{equation}
	\varkappa_c=-\varepsilon\frac{ K_0^\varepsilon a^2}{4C_d}.
\end{equation}
Below we show how this transition manifests itself in the density of states near the edge of the band
and the position of the mobility threshold.

{\it Solution of the RG equations.}
The RG flow equations (\ref{lambdaflow0}) and (\ref{gammaflow0}) [(\ref{lambdaflow}) and (\ref{gammaflow})]
can be solved exactly\cite{Syzranov:Weyl} with the result
\begin{align}
	&\tilde\varkappa(K)=
	\frac{\varkappa}{1-\frac{\varkappa}{\varkappa_c}+\frac{\varkappa}{\varkappa_c}\left(\frac{K_0}{K}\right)^\varepsilon},
	\label{gammaanswer}
	\\
	&\lambda(K)=[{\tilde\varkappa(K)}/{\varkappa}]^{1/4}.
	\label{lambdasolution}
\end{align}

If the renormalisation procedure is terminated at weak disorder, $\gamma\ll1$, the
action (\ref{Action1})-(\ref{Action3}) with renormalised parameters (\ref{gammaanswer})-(\ref{lambdasolution})
can be used to compute low-energy physical observables, such as
conductivity\cite{Syzranov:Weyl} and the density of states, evaluated in the next section.

%%%%%%%%%%%%%%%%%%%%%%%%%%%%%%%%%%%%%%%%%%%%%%%%%%%%%%%%%%%%%%%%%%%%%%%%%%%%%%%%%%%%%%%
\section{Density of states and mobility threshold}
\label{Sec:DoS}

\subsection{Scaling analysis for the density of states}
\label{Subsec:scaling}

The existence of the critical point {in a semiconductor [material with quasiparticle dispersion
(\ref{eq:spectrum}) in the orthogonal symmetry class]} 
in $d>2 \alpha$
dimensions suggests that the density of states exhibits a critical behaviour near this point. 
Such behaviour 
is dramatically different from the conventional case of low dimensions ($d<2\alpha$), where the critical point
is absent and the disorder-averaged density of states and mobility threshold
is known to be a smooth function
of the disorder strength\cite{Bulka,Efetov:book,GarciaGarcia}.

In what immediately follows we use general scaling arguments to describe the density of states
near the critical point.
In the next subsections we confirm this critical behaviour by a microscopic
calculation in the limit of small $\varepsilon$.

According to the conventional phenomenology,
near a continuous transition one expects the existence of a single dominant correlation length scale
\begin{equation}
	\xi(\varkappa,E)=E^{-\frac{1}{z}} g\left[(\varkappa-\varkappa_c)/E^{\frac{1}{z\nu}}\right],
	\label{CorrrLength}
\end{equation}
where $\nu$ and $z$ are the correlation-length and dynamical critical exponents respectively,
and the energy $E$ is measured from the renormalised edge of the band\footnotemark[\value{footnote}].
For small energies [$E\ll E^*=c(\varkappa-\varkappa_c)^{z\nu}$] and supecritical disorder 
($\varkappa>\varkappa_c$)
it diverges as 
\begin{equation}
	\xi(\varkappa) \propto |\varkappa - \varkappa_c|^{-\nu}.
	\label{XiKappa}
\end{equation}
We note, that, in contrast, the transition across a non-zero energy $E^*$ (mobility threshold) is
described
by the conventional-Anderson-transition critical behaviour, where a distinct
localisation length $\xi_{\text{loc}}\propto|E-E^*(\varkappa)|^{-\nu_A}$ diverges,
while the correlation length $\xi$ remains finite,
see
Fig.~\ref{Fig:MobEdge}.
Finally, sufficiently close to the critical point ($\varkappa\approx\varkappa_c$)
\begin{equation}
	\xi\propto E^{-\frac{1}{z}}.
	\label{XiE}
\end{equation}

Following the conventional paradigm\cite{Sachdev:Book},
near critical point physical quantities are expected to be expressible in terms of
this divergent correlation length. 
According to this, we expect the density of states to have the scaling dimensions
of density over energy, $\rho \sim \xi^{-d}/E$, and thus to exhibit the form
\begin{equation}
	\rho(E,\varkappa)=E^{\frac{d}{z}-1}\Phi\left[(\varkappa-\varkappa_c)/E^{\frac{1}{z\nu}}\right]
	+\rho_{\text{smooth}},
	\label{ScalingGeneral}
\end{equation}
where $\Phi(x)$ is a universal scaling function.
Here $\rho_{\text{smooth}}$ is an analytic contribution to the density of states in the conduction band,
derived from the same rare-regions effects as the Lifshitz tail.
In what follows we consider the states in the conduction band and neglect the latter non-perturbative
instantonic
contribution, Eq.~(\ref{HighDTail}),
since it is suppressed by sufficiently large energy $E$ and small $\varepsilon$.

Based on Eq.~(\ref{ScalingGeneral}) we expect
that close to the critical disorder strength, $\varkappa\approx\varkappa_c$, $\Phi(x\rightarrow0)\rightarrow const$ and
the density of states near the edge of the band depends on the energy as
\begin{equation}
	\rho(E)\propto E^{\frac{d}{z}-1}.
	\label{RhoScalCrit}
\end{equation}

If the disorder is stronger than critical, $\varkappa>\varkappa_c$, the states with sufficiently
small energies are localised, and their density is smeared by disorder. Requiring that the density 
of states is energy-independent
dictates that in this limit $\Phi(x) 
  \rightarrow   x^{z \nu(d/z -1)}$, leading to a prediction of
\begin{equation}
	\rho_{strong}\propto(\varkappa-\varkappa_c)^{(d-z)\nu}.
	\label{RhoScalStr}
\end{equation}

For subcritical disorder, $\varkappa<\varkappa_c$ the dimensionless disorder strength flows to smaller
values under the RG, leading the absence of localisation in the conduction band (provided $d>2$
in addition to $d>2\alpha$). The density of states vanishes when approaching the (renormalised) edge
of the band (until
the Lifshitz tail is reached), but may depend on the strength of disorder.  Assuming that the disorder $\tilde\varkappa(K)$
strength and the parameter $\lambda(K)$ in the renormalised Lagrangian (\ref{Action2}) saturate at
constant values as $K\rightarrow0$ [as is also supported by the microscopic RG analysis, cf.
Eqs.~(\ref{gammaanswer}) and (\ref{lambdasolution})],
we expect the resulting energy dependence of the density of states to be given by the disorder-free
expression $\propto E^\frac{d-\alpha}{\alpha}$.
This requires that the scaling function in this regime
has the form $\Phi(x) \propto |x|^{-d\nu(z/\alpha-1)}$, which from Eq.~(\ref{ScalingGeneral}) then gives
\begin{equation}
	\rho(E,\varkappa)\propto(\varkappa_c-\varkappa)^{-d\nu\left(\frac{z}{\alpha}-1\right)}
	E^{\frac{d-\alpha}{\alpha}}.
	\label{RhoScalWeak}
\end{equation}
In this regime, the density of states thus exhibits a universal prefactor, that singularly
enhances the disorder-free density of states, diverging
as the transition at $\varkappa = \varkappa_c$ is approached from below.

The three regimes, described by Eqs.~(\ref{RhoScalCrit})-(\ref{RhoScalWeak}),
are summarised in Fig.~\ref{Fig:CritDos}.

%%%%%%%%%%%%%%%%%%%%%%%%%%%%%%%%%%%%%%%%%%%%%%%%%%%%%%%%%%%%%%%%%%%%%%%%%%%%%%%%
\subsection{Scaling analysis for the mobility threshold}

In the previous section and in Subsection~\ref{Subsec:scaling}, using scaling and a detailed  RG analysis,
we have found that  for $d > d_c$ and subcritical disorder 
strength $\varkappa < \varkappa_c$,
the effective disorder strength
vanishes at low energies, and all states in the conduction band remain extended.
Thus, for $\varkappa < \varkappa_c$ we expect the mobility threshold to be stuck inside 
or just above the Lifshitz tail, and in the $\varepsilon\rightarrow0$ limit pinned to the
bottom\footnotemark[\value{footnote}]
of the conduction band. 

In contrast,
if the disorder is stronger than critical, the disorder strength flows to larger
values, leading to the localisation of states with sufficiently small energies. If the energy
is not sufficiently small, the RG flow may be terminated
while the disorder is still weak, leading to the absence of localisation.

Thus, for $\varkappa>\varkappa_c$ we predict the existence of
a finite mobility threshold $E^*(\varkappa)$
in the conduction band
that separates localised and delocalised states. 
According to the scaling theory, 
we predict the mobility threshold, $E^*\propto\xi^{-z}$, to have the universal scaling form
\begin{equation}
	E^*(\varkappa)\propto\left(\varkappa-\varkappa_c\right)^{z\nu}.
	\label{MobThreshScaling}
\end{equation}

According to the scaling hypothesis,
the energy scale $E^*$, Eq.~(\ref{MobThreshScaling}), also happens to be the 
characteristic energy scale at which
the high-energy density of states (\ref{RhoScalCrit}) for $\varkappa<\varkappa_c$
crosses over to the density of states (\ref{RhoScalWeak}) in the effective disorder-free regime,
see Fig.~\ref{Fig:CritDos}.

\subsection{Scaling analysis for the localisation length}

%We can discuss the scaling behavior of the localization length with similar arguments. 

We first note that the correlation length $\xi(E,\varkappa)$, Eq.~(\ref{CorrrLength}),
of the state with energy $E$ for disorder strength $\varkappa$
in general should be contrasted with the localisation length $\xi_{\text{loc}}(E,\varkappa)$
near the Anderson
transition [near the mobility threshold $E=E^*(\varkappa)$], studied
in this subsection.

Because for $\varkappa=\varkappa_c$ the Anderson transition occurs at zero energy,
%Close to the critical point ($\varkappa=\varkappa_c$, $E=0$), 
the two localisation lengths,
$\xi_{loc}$ and $\xi$, are proportional to each other near the critical point ($\varkappa=\varkappa_c$, $E=0$).
This allows us to develop a scaling theory, similar to that
of Subsection~\ref{Subsec:scaling}, for the localisation length 
\begin{equation}
\xi_{\rm loc}(E,\varkappa) =(\varkappa-\varkappa_c)^{-\nu}
h \left[(\varkappa-\varkappa_c)/E^{\frac{1}{z\nu}}\right],
\end{equation}
where $h(x)$ is a universal scaling function.
%where $h(x)$ is a scaling function, distinct from $g(x)$ in Eq.~(\ref{CorrrLength}).

Close to the critical point ($\varkappa=\varkappa_c$, $E=0$) the scaling of the localisation length
for sufficiently-low-energy states [$E\ll c(\varkappa-\varkappa_c)^{-\nu}$] is thus given 
by Eq.~(\ref{XiKappa}) for disorder close to critical.
However, for $\varkappa>\varkappa_c$ and as $E\rightarrow E^*(\varkappa)$
the critical behaviour of the localisation length is of
the Anderson-transition universality class [see Eq.~(\ref{XiAnderson})],
with the correlation length $\xi$ remaining finite.
This dictates the following $\varkappa\geq\varkappa_c$ form of the localisation length: 
\begin{align} 
\xi_{\rm loc} (\varkappa, E) \propto  \left(\varkappa-\varkappa_c \right)^{-\nu} 
\left[ \frac {E}{\left( \varkappa- \varkappa_c \right)^{z \nu} }- c
\right]^{-\nu_A} 
\nonumber\\
=  
\left(\varkappa-\varkappa_c\right)^{\nu\left(z\nu_A - 1 \right) }
\left[E - c\left( \varkappa-\varkappa_c \right)^{z \nu} \right]^{-\nu_A}, \label{eq:loc4}
\end{align}
where $\nu_A$ is the correlation-length exponent of the Anderson transition.
Eq.~(\ref{eq:loc4}) holds for energies 
in the vicinity of the mobility threshold $E^*(\varkappa)=c(\varkappa-\varkappa_c)^{z\nu}$,
within the blue (grey) wedge-shaped region in Fig.~\ref{Fig:MobEdge}.

We emphasise that the divergence of the localisation length is characterised by different critical exponents
at the (high-dimensional)
critical point $\varkappa=\varkappa_c$, $E=0$ and at $\varkappa>\varkappa_c$, $E=E^*(\varkappa)$.
Indeed, at the former the
correlation-length and dynamical exponents are given by $\nu$ and $z$ respectively, while for $\varkappa>\varkappa_c$--
by $\nu_A$ and\cite{ShapiroAbrahams:z,Wegner:z} {$z_A=d$}.

The localisation transition for subcritical disorder ($\varkappa<\varkappa_c$) occurs in a narrow
interval of energies close to the bottom of the band, where the states cross over to the Lifshitz tail.
Expecting that the nature of such transition is thus affected by rare-regions strong-disorder
effects, we leave it for future studies.

In what follows we 
complement the above scaling analysis by a microscopic derivation of 
Eqs.~(\ref{RhoScalCrit})-(\ref{MobThreshScaling}) and compute the critical
exponents $\nu$ and $z$ and the associated scaling functions microscopically in the
limit of small $\varepsilon=d-2\alpha<0$.

%%%%%%%%%%%%%%%%%%%%%%%%%%%%%%%%%%%%%%%%%%%%%%%%%%%%%%%%%%%%%%%%%%%%%%%%%%%%%%%%%%%%%%%%%%%%%%%%%%%%%%%%

%\subsection{General expression for the density of states}

\subsection{Microscopic calculation of the density of states and mobility threshold in high dimensions, $d>2\alpha$}

In the absence of disorder the density of states in the conduction band is given by
\begin{equation}
	\rho_{clean}(E)=\frac{C_d E^\frac{d-\alpha}{\alpha}}{\alpha a^\frac{d}{\alpha}},
	\label{RhoClean}
\end{equation}
and the Lifshitz tail is absent.

In the presence of disorder,
the density of states $\rho(E)$ can be calculated microscopically from the renormalised 
Lagrangian, Eqs.~(\ref{Action1})-(\ref{Action3}) with the cutoff $K_E$,
determined by Eq.~(\ref{MomTerminate}).
Provided the renormalised disorder remains weak, $\gamma(K_E) \ll 1$, this can be done in a controlled perturbative expansion in  $\gamma(K_E)$, with the lowest-order contribution given simply by the quadratic part of the Lagrangian,
Eq.~(\ref{Action2}), utilising the renormalised parameters $\varkappa(K_E)$ and $\lambda(K_E)$, Eqs.~(\ref{gammaanswer})
and (\ref{lambdasolution}). 

To this leading order in $\gamma(K_E)$, we thus find
\begin{subequations}
\begin{align}
	\rho(E,\varkappa)
	&= 
	\lambda(K_E)\cdot\rho_{clean}\left[\lambda(K_E)E\right]
	\\
	&=\frac{C_d }{\alpha a^\frac{d}{\alpha}}\left[\lambda(K_E)\right]^\frac{d}{\alpha}E^\frac{d-\alpha}{\alpha}
	\label{RhoUnivLambda}
	\\
	&=\frac{C_d }{\alpha a^\frac{d}{\alpha}}
	\left[1-\frac{\varkappa}{\varkappa_c}+\frac{\varkappa}{\varkappa_c}
	\left(\frac{K_0}{K_E}\right)^\varepsilon\right]^{-\frac{d}{4\alpha}}E^\frac{d-\alpha}{\alpha},
	\label{RhoUniv}
\end{align}
\end{subequations}
where the momentum $K_E$, at which the RG flow is terminated, is a function of energy $E$,
determined by the condition
\begin{equation}
	E\left[1-\frac{\varkappa}{\varkappa_c}+\frac{\varkappa}{\varkappa_c}
	\left(\frac{K_0}{K_E}\right)^\varepsilon\right]^{-\frac{1}{4}}\sim a K_E^\alpha,
	\label{CutoffECondition}
\end{equation}
as follows from Eqs.~(\ref{MomTerminate}), (\ref{gammaanswer}), and (\ref{lambdasolution}).

Because the disorder strength $\tilde\varkappa(K)$ always increases under the RG flow, according
to Eq.~(\ref{gammaflow0}), the parameter $\lambda(K)$ is always larger than unity.
Therefore,
the low-energy density of states (\ref{RhoUnivLambda})
in a disordered system exceeds that (\ref{RhoClean})
in a disorder-free system.
Thus, impurities have transferred states from high energies $E>aK_0^\alpha$ to
lower energies.

Examining Eqs.~(\ref{RhoUniv}) and (\ref{CutoffECondition}) it is clear, that the density of states exhibits
three qualitatively different regimes, distinguished by the range of the momentum
cutoff $K_E$ (or correspondingly energy $E$) and on whether the disorder is
stronger or weaker than critical.

Indeed, comparing the terms $1-\varkappa/\varkappa_c$
and $(\varkappa/\varkappa_c)(K_0/K_E)^\varepsilon$ in Eqs.~(\ref{RhoUniv}), (\ref{CutoffECondition}),
(\ref{gammaanswer}), and (\ref{lambdasolution}) suggests an introduction of the momentum  
scale
\begin{equation}
	K^*=K_0\left|1-\frac{\varkappa_c}{\varkappa}\right|^{-\frac{1}{\varepsilon}}
	\label{Kmob}
\end{equation}
and the corresponding energy scale $E^* = a (K^*)^\alpha/\lambda(K^*)$ given by
\begin{equation}
	E^*=aK_0^\alpha\left|1-\frac{\varkappa_c}{\varkappa}\right|^{\frac{1}{4}-\frac{\alpha}{\varepsilon}}.
	\label{CritEnergy}
\end{equation}
The three regimes are defined by the energy $E$ and disorder strength:
(1) disorder close to critical, $\varkappa\approx\varkappa_c$, corresponding to the
energy range, such that $K^*\ll K_E<K_0$;
(2) subcritical disorder and low energies, $\varkappa<\varkappa_c$ and $K_E\ll K^*$;
(3) supercritical disorder and low energies, $\varkappa>\varkappa_c$ and $K_E\ll K^*$.

The analysis of whether corresponding energy-$E$ states
are %a state with energy $E$ is 
localised can be carried out similarly
to the case of a usual metal\cite{Efetov:book}. In $d\leq 2$ dimensions all the states are localised.
In the dimensions $d>2$ there is a mobility threshold $E^*$, corresponding
to $K_{E^*}\ell(K_{E^*})\sim 1$ [$\gamma(K_{E^*})\sim 1$],
that separates localised and delocalised states.

In what immediately follows, we compute the density of states in these three regimes. 

%%%%%%%%%%%%%%%%%%%%%%%%%%%%%%%%%%%%%%%%%%%%%%%%%%%%%%%%%%%%%%%%%%%%%
\subsubsection{Critical disorder}

In the case of disorder close to critical, $\varkappa\approx\varkappa_c$,
corresponding to the interval of energies $E^*\ll E<aK_0^\alpha$,
Eq.~(6.11), relating
the momentum $K_E$ to the energy $E$, simplifies to
\begin{equation}
	K_E=K_0\left(\frac{E}{aK_0^\alpha}\right)^\frac{4}{4\alpha-\varepsilon}.
\end{equation}
This, together with
Eq.~(\ref{RhoUniv}), yields the critical density of states
in this energy interval 
\begin{equation}
	\rho(E)\sim\frac{C_dK_0^{d-\alpha}}{\alpha a}
	\left(\frac{E}{aK_0^\alpha}\right)^\frac{3d-2\alpha}{2\alpha+d}.
	\label{SubLowTop}
\end{equation}

For energies of the order of or larger than the ultra-violet cutoff,
$E\gtrsim aK_0^\alpha$, the density of states crosses over to that of a clean semiconductor, Eq.~(\ref{RhoClean}).

%%%%%%%%%%%%%%%%%%%%%%%%%%%%%%%%%%%
\subsubsection{Subcritical disorder}

In this regime of $\varkappa < \varkappa_c$, defined by
$K_E < K^*$, the system is sufficiently away from the critical disorder strength $\varkappa_c$,
so that $\frac{\varkappa}{\varkappa_c}
	\left(\frac{K_0}{K_E}\right)^\varepsilon$ in Eqs.~(\ref{RhoUniv}) and (\ref{CutoffECondition}) can be neglected in comparison
with $1-\varkappa/\varkappa_c$. Equation (\ref{RhoUniv}) then immediately gives
\begin{equation}
	\rho(E)=\frac{C_d}{\alpha a^\frac{d}{\alpha}}
	\left(1-\frac{\varkappa}{\varkappa_c}\right)^{-\frac{d}{4\alpha}}
	E^\frac{d-\alpha}{\alpha},
	\label{SubLowBot}
\end{equation}
a result that applies for subscritical disorder and sufficiently low energies $E \ll E^*$,
as illustrated in Fig.~\ref{Fig:CritDos}.
The disorder-averaged low-energy density of states is asymptotically that of a disorder-free
semiconductor, with the only effect of the random potential to enhance the density of states
through a universal multiplicative prefactor, that diverges near the critical point.

For weak disorder, $\varkappa\ll\varkappa_c$ the renormalisation is weak, 
and the density of states (\ref{SubLowBot}) is close to that (\ref{RhoClean})
of a clean semiconductor.

%%%%%%%%%%%%%%%%%%%%%%%%%%%%%%%%%%%%%%%%%%%%%%
\subsubsection{Supercritical disorder}
\label{Subsubsec:supercritical}

For disorder stronger than critical, $\varkappa>\varkappa_c$,
the dimensionless measure of disorder $\gamma(K)\sim [k\ell(k)]^{-1}$ grows upon renormalisation.
It reaches values of order unity at
momentum cutoff $K_E\sim K^*$,
below which our perturbative (in $\gamma$) RG is no longer trustworthy.

However, one can apply phenomenological arguments of Subsection~\ref{Subsec:phenom} with
the renormalised strength of disorder $\varkappa^*\sim a^2\left(K^*\right)^\varepsilon$
at the RG breakdown point ($\gamma\sim 1$).
At this point, the root mean square 
$U_{\text{rms}}^*\sim\left[\varkappa^*\left(K^*\right)^d\right]^\frac{1}{2}$
of the renormalised random potential is comparable to
the kinetic energy $a\left(K^*\right)^\alpha$.
Therefore, we expect that the states with energy $E<E^*$, where $E^*$ is given by Eq.~(\ref{CritEnergy}),
are strongly influenced by such strong random potential and are thus localised.
Conversely, for $E > E^*$ and $d>2$ the random potential is a small perturbation and the states are delocalised.

Given that $\gamma \sim [k\ell(k)]^{-1}$ [see Eq.~(\ref{GammaDefinition})],
this conclusion is also consistent with the Ioffe-Regel criterion of localisation
(supported by rigorous analytic calculations\cite{Efetov:book,Kamenev:book}).

We thus conclude that for $d > d_c$ the energy scale $E^*(\varkappa)$, Eq.~(\ref{CritEnergy}),
defines the mobility threshold for the strong disorder regime $\varkappa > \varkappa_c$,
separating localised and delocalised states (so long as $d>2$), as illustrated
in Figs.~\ref{Fig:CritDos} and \ref{Fig:MobEdge}.

Because the
disorder is strong for states with energies $E<E^*$,
the density of states is energy-independent and is
determined by the amplitude of the disorder potential fluctuations. 

From Eqs.~(\ref{RhoUniv}) and (\ref{Kmob})
we obtain the density of states for $\varkappa>\varkappa_c$ in the
energy interval $0<E\lesssim E^*$:
\begin{align}
	\rho(E)
	\sim\frac{C_d K_0^{d-\alpha}}{\alpha a}
	\left(\frac{\varkappa-\varkappa_c}{\varkappa}\right)^{\frac{3}{4}-\frac{\alpha}{\varepsilon}}.
\end{align}

\subsection{Subcritical dimensions, $d<2\alpha$}

Below critical dimensions ($\varepsilon>0$),
the disorder strength grows
with decreasing energy $E$, appearing to diverge as $K_E$ approaches $K^*$.
The dimensionless measure of disorder, $\gamma(K)$ reaches values of order unity at
momentum cutoff $K_{\text{mob}}= K^*[1+|\varkappa_c|/\tilde\varkappa(K)]^{1/\varepsilon}\sim K^*$,
below which our perturbative (in $\gamma$) RG is no longer trustworthy.

Similarly to the case of supercritical disorder in higher dimensions, the 
momentum $K_{\text{mob}}$ corresponds to the mobility threshold $E_{\text{mob}}$,
if $d>2$ (in addition to $d<2\alpha$), all states with $E<E_{\text{mob}}$ being localised.

Using the condition $\gamma(K)\sim1$ and Eqs.~(\ref{MomTerminate}), (\ref{lambdasolution}), and (\ref{Kmob}),
we obtain the mobility threshold in such lower dimensions:
\begin{equation}
	E_{\text{mob}}=aK_0^\alpha
	\left(\frac{\varkappa}{a^2K_0^\varepsilon}\right)^\frac{1}{4}
	\left(1+\frac{|\varkappa_c|}{\varkappa}\right)^{\frac{1}{4}-\frac{\alpha}{\varepsilon}}.
	\label{SubMobExact}
\end{equation}

Finally, we note, that sufficiently below critical dimension, $\varepsilon\gtrsim1$, Eq.~(\ref{gammaanswer}) shows that,
in agreement with the perturbation theory of Sec.~\ref{Sec:Perturbation},
the renormalisation of the disorder strength is negligible, i.e. $\tilde\varkappa(K)\approx\varkappa$
so long as the disorder is weak, $\gamma\ll1$.
In contrast, just below the critical
dimension ($0<\varepsilon\ll1$) the parameters of the system are significantly renormalised
due to elastic scattering between states in the whole conduction band.

\section{Weyl semimetal}
\label{Sec:Weyl}

Weyl semimetal is a 3D material
characterised by Dirac quasiparticle dispersion of long-wave excitations,
\begin{equation}
	\cH=v\hbsigma\cdot\bk,
	\label{DiracDispersion}
\end{equation}
with $\hbsigma$ being a (pseudo)spin-1/2 operator.

Generically one expects an even number of Dirac points in the first Brillouin zone
(a consequence of Dirac fermion doubling problem on a lattice\cite{Nielsen:doubling}).
However, for sufficiently smooth random potential, that we will assume here for simplicity,
scattering between Dirac points (internodal scattering) may be neglected, restricting the analysis to the
vicinity of one point only.

We first note, that, unlike the case of a semiconductor described by the model (\ref{Hamiltonian}),
quasiparticles in Weyl semimetal cannot be localised in the absence of internodal scattering.
This follows from the observation that Weyl fermion is characterised by a non-zero Berry flux through a closed surface
surrounding the Dirac point in the momentum space\cite{Wan:WeylProp}. Thus, WSM may be considered
as a surface of a 4D topological insulator in the AII class\cite{RyuLudwig:classification}.
Surface states of a topological insulator cannot get localised by disorder, and, thus, neither can Weyl fermions
near one Dirac point.

Despite the absence of the Anderson transition in Weyl semimetal,
a weak-to-strong disorder transition manifests itself in a critical behaviour
of a variety of physical observables, in particular the disorder-averaged density of states,
to whose analysis we now turn.

The RG analysis for disordered materials with Dirac-type quasiparticle dispersion
is similar to that for high-dimensional semiconductors, described in Section~\ref{Sec:RG},
and have been carried out in a number of previous
works\cite{DotsenkoDotsenko,Fradkin1,Fradkin2,LudwigFisher,Nersesyan:dwave,Goswami:TIRG,AleinerEfetov,Syzranov:Weyl,Moon:RG}.

The critical dimension in the case of quasiparticles with linear dispersion, Eq.~(\ref{DiracDispersion}),
is $d_c=2\alpha=2$, and thus the RG treatment of disorder and the aforementioned
weak-to-strong disorder transition (unlike, conventional semiconductors
studied in earlier sections) is of direct physical relevance in 3D Dirac materials, WSM.

In order to have a ``controlled'' RG calculation in WSM,
it is essential to analytically continue the model to an arbitrary dimension $d$
and then perform an $\varepsilon=2-d$-expansion.
We do this by analytically continuing the quasiparticle dispersion according to
\begin{equation}
	\cH=vk^{\frac{1}{2}+\frac{\varepsilon}{2}}\hbsigma\cdot\bk,
	\label{DiracDispersionEpsilon}
\end{equation}
and setting $\varepsilon = -1$ at the end of the calculation.

{{Perturbative}} RG analysis, quite similar to that of Sec~\ref{Sec:RG},
together with such $\varepsilon$-expansion\cite{Fradkin1,Goswami:TIRG,Syzranov:Weyl} leads
{{in the one-loop approximation}}
to the flow equations
\begin{align}
	\partial_l\lambda &=\gamma\lambda/2,
	\label{lambdaflow1}
	\\
	\partial_l\gamma &=\varepsilon\gamma+\gamma^2,
	\label{gammaflow1}
\end{align}
that have the same form as Eqs.~(\ref{lambdaflow}) and (\ref{gammaflow}),
except for a different prefactor, $1/4\rightarrow 1/2$,
in Eq.~(\ref{lambdaflow1}) and in the definition
of the dimensionless disorder strength
\begin{equation}
	\gamma(K)=\frac{2C_d}{v^2}\tilde\varkappa(K)K^{-\varepsilon}.
	\label{GammaDefinitionW}
\end{equation}

Following the scheme of Sec.~\ref{Sec:RG},
we immediately find the critical exponents in 3D (i.e. for WSM),
\begin{equation}
	\nu=1,\quad z=\frac{3}{2},
	\label{ExponentsWeyl}
\end{equation}
and the critical disorder strength
\begin{equation}
	\varkappa_c=\pi^2v^2/K_0,
\end{equation}
which have also been obtained in
the previous works\cite{Fradkin1,Goswami:TIRG,Syzranov:Weyl,Moon:RG}.
{The values of the critical exponents close to (\ref{ExponentsWeyl}) have also been found numerically
in Ref.~\onlinecite{Herbut}.}

Although localisation in Weyl semimetal is forbidden by symmetry in the absence 
of internodal scattering, the 
disorder-driven phase transition manifests itself
in the conductivity and the density of states.
The conductivity of Weyl semimetal for small finite doping
has been calculated microscopically in Ref.~\onlinecite{Syzranov:Weyl}.

The density of states in Weyl semimetal can be evaluated similarly to that of a high-dimensional
semiconductor, described in detail in Sec.~\ref{Sec:DoS},
by solving the above flow equations (\ref{lambdaflow1}) and (\ref{gammaflow1})
and 
using the quadratic part of the
quasiparticle Lagrangian with renormalised couplings, 
which is justified for weak renormalised disorder $\gamma(K_E) \ll 1$.

In the absence of disorder,
\begin{equation}
	\rho_{clean}^{Weyl}(E)=\frac{E^2}{2\pi^2v^3}.
\end{equation}

For disorder
strength $\varkappa$ close to the critical $\varkappa_c$, the RG analysis yields
\begin{equation}
	\rho(E)\sim\frac{K_0}{v^2}E
	\label{RhoWeylClseCrit}
\end{equation}
in the energy interval $E^*_{Weyl}\ll E< vK_0$,
where
\begin{equation}
	E^*_{Weyl}=vK_0\left|1-\frac{\varkappa_c}{\varkappa}\right|^{\frac{3}{2}}
\end{equation}
is the crossover energy scale that for $\varkappa < \varkappa_c$ delineates linear (critical, $E > E^*$)
and quadratic (disorder-free, $E < E^*$) behaviour of the density of states.

For subcritical disorder, $\varkappa<\varkappa_c$, and energies $0<E\ll E^*_{Weyl}$,
the flows (\ref{lambdaflow1}) and (\ref{gammaflow1})
crossover from the vicinity of the critical point to the disorder-free Gaussian fixed point.
In this regime we find the disorder-free $E^2$ scaling of the density of states as a function of energy, 
enhanced by a universal singular prefactor that diverges as $\varkappa$ approaches the critical value:
\begin{equation}
	\rho(E)=\frac{1}{2\pi^2v^3}\left(1-\frac{\varkappa}{\varkappa_c}\right)^{-\frac{3}{2}}
	E^2.
\end{equation}

Strong random potential, $\varkappa>\varkappa_c$, is relevant, leading to
the density of states smeared by disorder
and independent of energy in the interval for $|E|\lesssim E^*_{Weyl}$:
\begin{equation}
	\rho(E)\sim\frac{K_0^2}{v}\left(1-\frac{\varkappa_c}{\varkappa}\right)^\frac{3}{2}.
	\label{RhoWeylWeak}
\end{equation}

The critical regimes, described by Eqs.~(\ref{RhoWeylClseCrit})-(\ref{RhoWeylWeak}),
are summarised in Fig.~\ref{Fig:CritDosWeyl}.

%%%%%%%%%%%%%%%%%%%%%%%%%%%%%5OUTLOOK%%%%%%%%%%%%%%%%%%%%%%%%%%%%%%%%%%%%%%%%%%%%%%%%%%%%%%%%%%%%%%

\section{Conclusion}
\label{Sec:outlook}

\subsection*{Summary}

In this work we have studied noninteracting quasiparticles with power-law dispersion 
moving in a weak random potential. We demonstrated that in contrast to low dimensions
(where for $2  < d < 2\alpha$, a conventional Anderson localisation transition takes place),
for $d > 2\alpha$ such system in addition exhibits a disorder-driven transition in a new universality class.
Among other physical properties, it manifests itself in a universal critical behaviour
of the disorder-averaged density of states and in
the sharp dependence of the mobility threshold on disorder strength $\varkappa$.
In particular, the mobility threshold vanishes for $\varkappa$ smaller than a critical value. 
These results are summarised by Figs.~\ref{PhaseDiagram}-\ref{Fig:DoSpictureWeyl}.

\subsection*{Outlook}

In light of our finding of a novel localisation transition and its phenomenology
in high dimensions, natural future research directions include its interplay with
interactions, more generic band structures (e.g., including other bands) and disorder symmetries,
spin-orbital coupling,
magnetic field, etc.

Another issue that our work raises is the nature of the high-dimensional localisation transition
for $\varkappa < \varkappa_c$, across the mobility edge, located close to the edge of
the conduction band.
Although one may expect that this transition is in the conventional Anderson-localisation class,
this question %certainly 
deserves further investigation.

Also, we suggest that a localisation transition on the Cayley tree, believed to correspond
to the infinite dimension $d=\infty$,
deserves further investigation, as it may realise the high-dimensional phenomenology studied here. 
Indeed,
it is well-known that including states with energies
far from the Fermi level is necessary
to describe localisation and transport on Cayley tree\cite{AbouThouAnd,MirlinFyodorov,Efetov:book},
similarly to the case of high-dimensional semiconductors considered in this paper.
We thus expect models on Cayley tree to display the striking phenomenology uncovered here,
leading to, e.g., a critical behaviour
of the disorder-averaged density of states or novel universality classes of the
localisation transition.

Another class of systems, which exhibit similar unconventional single-particle interference effects,
that involve elastic scattering between all states in the band,
is lattice models with strong on-site disorder and weak inter-site hopping\cite{Syzranov:accond,Syzranov:MFRG},
describing, e.g, strongly disordered insulators or granulated superconductors in the insulating states.
Because such systems can be analysed by means of a similar RG approach,
with momentum states replaced by (quasi-)localised on-site states,
we expect superconductor-insulator transitions and metal-insulator transitions in such systems to display
similar phenomenology.

{\it Acknowledgements.} We have benefited from discussions with B.L.~Altshuler, V. Dobrosavljevi{\'c},
M.P.A.~Fisher,
A.~Kamenev, B.I.~Shklovskii, and C.~Tian.
Our work has been supported by the Alexander von Humboldt
Foundation through the Feodor Lynen Research Fellowship (SVS) and by the NSF grants
DMR-1001240 (LR and SVS),
DMR-1205303 (VG and SVS), PHY-1211914 (VG and SVS), and PHY-1125844 (SVS).
LR also acknowledges The Kavli Institute for Theoretical Physics, where a part
of this work was done, for its support through the National Science Foundation
under the grant PHY11-25915, as well as a partial support by the Simons
Investigator award from the Simons Foundation.

\appendix

\section{Mean free path in a weakly disordered semiconductor}
\label{Sec:MFP}

In this section we compute the mean free path for
a quasiparticle with energy $E$ 
in a $d$-dimensional semiconductor with short-range-correlated disorder.

Let us consider first the case of very weak disorder,
$\varkappa\ll\varkappa_c$. Then 
the mean free path is dominated by elastic scattering 
in a narrow momentum shell around the surface $ap^\alpha=E$ in momentum space.
The renormalisations due to the elastic scattering through the states
far from this surface can be neglected.

The self-energy part can then be found by integrating over 
momentum states $\bp$ with
energies close to $E$:
\begin{equation}
	\Sigma^R(E)=-\varkappa \:\int G^R(\bp,E)\frac{d\bp}{(2\pi)^d}
	\label{SigmaR}
\end{equation}
with the bare Green's function
\begin{equation}
	G^R(\bp,E)=\left(E-ap^\alpha+i0\right)^{-1}.
	\label{GreenFree}
\end{equation}

The mean free path is then given by
\begin{equation}
	\ell(k)=-\frac{v(k)}{2\:\Im\:\Sigma^R(E)},
	\label{MeanFreeGen}
\end{equation}
where $E=ak^a$ and $v(k)=\alpha ak^{\alpha-1}$ is the velocity corresponding to the momentum $k$.
Using Eqs.~(\ref{SigmaR})-(\ref{MeanFreeGen}) we immediately arrive at the result (\ref{MeanFreePath})
for the mean free path in a weakly disordered semiconductor.

If disorder is not very weak ($\varkappa\sim\varkappa_c$ or $\varkappa>\varkappa_c$),
quasiparticle properties experience renormalisation from elastic scattering
between all states in the band. By applying the RG procedure, described in Sec.~\ref{Sec:RG}, it is possible
to remove high momenta from the system and reduce the problem to considering only momentum
states with energies close to $E$. 

In particular, if disorder is not very weak but still smaller than critical, $\varkappa<\varkappa_c$,
the system flows towards vanishing disorder, and the elastic scattering rate in the renormalised system
can be obtained in the Born approximation similarly to the case of a usual low-dimensional metal
or a semiconductor\cite{AGD}.
The renormalised disorder strength can 
be sufficiently small for applying the Born approximation also in the case $\varkappa>\varkappa_c$
and small $\varepsilon = 2\alpha-d \ll 1$, if the 
RG procedure is terminated by sufficiently large energy $E$, while the disorder is still weak.

The self-energy part for a quasiparticle with energy $E$ then
is given by Eq.~(\ref{SigmaR}) with the replacement $E\rightarrow \lambda(k) E$ 
inside the argument of the Green's function (\ref{GreenFree}) on the right-hand side,
\begin{equation}
	\lambda(k) E=ak^\alpha.
\end{equation}

The respective mean free path can be also defined by Eq.~(\ref{MeanFreeGen}),
leading to Eq.~(\ref{MeanFreePath}). Then the small parameter $\gamma(k)\sim[k\ell(k)]^{-1}\ll1$
plays the same role in the renormalised system as it does in a usual metal\cite{AGD}
or a very-weakly-disordered non-renormalised system; it suppresses diagrams with
crossed impurity lines.

Indeed, single-particle interference effects involve quasiparticle propagators with equal energies $E$.
The suppression of diagrams with crossed impurity lines occurs due to an additional constraint\cite{AGD}
of the form
$|ap_1^\alpha\pm ap_2^\alpha\pm\ldots \pm ap_N^\alpha|\lesssim-\Im\:\Sigma^R(E)$
on the quasiparticle momenta $p_1, p_2, \ldots, p_N$ near the surfaces $ap_1^\alpha=ap_1^\alpha
=\ldots=ap_N^\alpha=\lambda(k)E$. Such diagrams are suppressed if $k\ell(k)\gg 1$ with the mean
free path $\ell(k)$ defined by Eq.~(\ref{MeanFreeGen}).

%%%%%%%%%%%%%%%%%%%%%%%%%%%%%%%%%%%%%%%%%%%%%%%%%%%%%%%%%%%%%%%%%%%%%%%%%%%%%%%%%%%%%%%%%%%%%%%%%%%%%%%%%%%%%%%
\section{Details of the RG analysis}
\label{Sec:RGDetail}

In this section we provide details of the renormalisation-group analysis
for the density of states (\ref{Action1}) and the
quasiparticle Lagrangian (\ref{Action2})-(\ref{Action2}).

On each step of the RG procedure we split the supervectors $\psi$, $\psi^\dagger$
into the ``fast'' $\psi_f$, $\psi_f^\dagger$ and ``slow'' $\psi_s$, $\psi_s^\dagger$ parts,
including respectively
the larger ($Ke^{-l}<k<K$) and the smaller ($k<Ke^{-l}$) momentum
components of the fields $\psi$, $\psi^\dagger$ and, perturbatively in 
the weak random potential, integrate out the fast components.

The Lagrangian of the quasiparticles 
separates into a sector containing only fast fields, a sector of the slow fields
and the ``interaction'' Lagrangian $\cL_i$ that couples fast and slow degrees of freedom: 
\begin{align}
	\cL(\psi^\dagger,\psi)
	=\cL(\psi_f^\dagger,\psi_f)+\cL(\psi_s^\dagger,\psi_s)
	\nonumber\\
	+\cL_i(\psi_s^\dagger,\psi_s,\psi_f^\dagger,\psi_f),
\end{align}
where $\cL(\psi^\dagger,\psi)=\cL_0(\psi^\dagger,\psi)+\cL_{int}(\psi^\dagger,\psi)$,
Eqs.~(\ref{Action2})-(\ref{Action3}), and
\begin{align}
	&\cL_i(\psi_s^\dagger,\psi_s,\psi_f^\dagger,\psi_f)
	\nonumber\\
	&=\tilde\varkappa\int(\psi_f^\dagger\psi_f)(\psi_s^\dagger\psi_s)d\br
	+\tilde\varkappa\int(\psi_s^\dagger\psi_f)(\psi_f^\dagger\psi_s)d\br
	\nonumber\\
	&+\frac{\tilde\varkappa}{2}\int(\psi_f^\dagger\psi_s)(\psi_f^\dagger\psi_s)d\br
	+\frac{\tilde\varkappa}{2}\int(\psi_s^\dagger\psi_f)(\psi_s^\dagger\psi_f)d\br.
	\label{LSlowFastInteractions}
\end{align}

Integrating out the fast field results in (i) the renormalisation of the Lagrangian
of the slow modes and (ii) the renormalisation
of the preexponential factor 
in the
expression (\ref{Action1}) for the density of states.

{\it Renormalised Lagrangian.} To the leading order in the small disorder strength
(one-loop approximation) the Lagrangian of the slow modes is renormalised according to
\begin{equation}
	\cL(\psi_s^\dagger,\psi_s)\rightarrow\cL(\psi_s^\dagger,\psi_s)
	+\langle\cL_i\rangle_f-\frac{1}{2}\llangle\cL_i^2\rrangle_f,
\end{equation}
where $\langle\ldots\rangle_f=\int \cD\psi_f^\dagger\cD\psi_f\ldots e^{-\cL(\psi_f^\dagger,\psi_f)}$,
and $\llangle\ldots\rrangle$ is a similar notation for irreducible (connected) correlators.

\begin{figure}[htbp]
	\centering
	\includegraphics[width=0.1\textwidth]{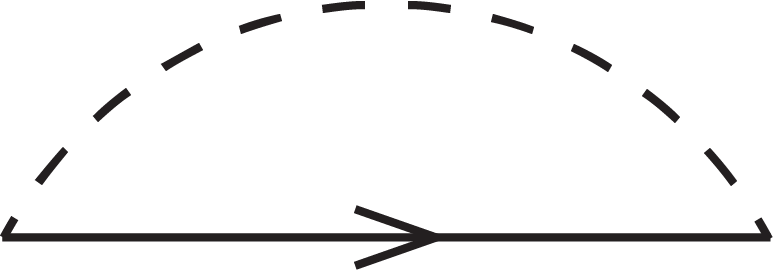}
	\caption{\label{Fig:SelfEnergy}
	Diagram corresponding to the renormalisation of the quadratic part of the Lagrangian.
	}
\end{figure}

The renormalisation of the quadratic part of the Lagrangian is determined by the term $\langle\cL_i\rangle_f$,
\begin{align}
	\delta\cL_0(\psi_s^\dagger,\psi_s)=\langle\cL_i\rangle_f
	=\tilde\varkappa\int\left<(\psi_s^\dagger\psi_f)(\psi_f^\dagger\psi_s)\right>_fd\br
	\nonumber\\
	=\int d\br\: \psi_s^\dagger(\br)\psi_s(\br)
	\cdot\tilde\varkappa
	\int\frac{d\bp}{(2\pi)^d}\frac{i}{E\cdot\lambda(K)-ap^\alpha+i0},
	\label{L2}
\end{align}
and, in terms of the disorder-averaging perturbation theory, corresponds to the diagram in Fig.~\ref{Fig:SelfEnergy}.

In deriving Eq.~(\ref{L2}) we used the correlator 
\begin{equation}
	\langle\psi_{f\bk}\psi_{f\bk}^\dagger\rangle_f=
	\holOne_{FB}\cdot
	\frac{i}{E\cdot\lambda(K)-ak^\alpha+i0},
\end{equation}
of the Fourier-transform of the supervectors 
$\psi^{(\dagger)}_{f\bk}=\frac{1}{\sqrt{V}}\int\psi^{(\dagger)}_{f}(\br)\exp(\mp i\bk\br)\:d\br$,
with $\holOne_{FB}$ being the unity matrix in the space of fermionic and bosonic components
of the supervectors.

The renormalisation (\ref{L2}) of the quadratic part of the Lagrangian leads to
a shift of the edge of the conduction band and a modification of the coupling $\lambda$:
\begin{align}
	\lambda(Ke^{-l})\cdot E\rightarrow \lambda(Ke^{-l})\cdot E + \lambda(Ke^{-l})\cdot \delta E+
	\delta\lambda\cdot E,
\end{align}
where
\begin{align}
	\delta E 
	&=\tilde\varkappa\int_{Ke^{-l}<p<K}\frac{d\bp}{(2\pi)^d}\frac{1}{ap^\alpha}
	\nonumber\\
	&=\frac{\tilde\varkappa C_dK^{d-\alpha}}{a(d-\alpha)}\left[1-e^{-(d-\alpha)l}\right]
\end{align}
describes the shift of the edge of the band. Throughout the paper we measure the energy $E$ from
the edge of the band, i.e., on each step of the RG procedure absorb $\delta E$ into the redefinition
of the energy $E$: $E+\delta E\rightarrow E$. The modification of the parameter $\lambda$ in
the limit of small $\varepsilon=2\alpha-d$ reads
\begin{align}
	\delta\lambda
	& =\tilde\varkappa\lambda\int_{Ke^{-l}<p<K}\frac{d\bp}{(2\pi)^d}\frac{1}{a^2p^{2\alpha}}
	\nonumber \\
	& \approx\frac{C_d}{a^2}\tilde\varkappa\lambda K^{-\varepsilon}
	\cdot l
\end{align}
and leads to the RG equation (\ref{lambdaflow0}).

The renormalisation of the disorder strength $\tilde\varkappa$
[of the quartic term in the Lagrangian, Eq.~(\ref{Action3})]
is described by irreducible (connected) pairwise correlators of different terms in the
right-hand-side of Eq.~(\ref{LSlowFastInteractions}) and corresponds
to the diagrams in Fig.~\ref{Fig:LineDiagrams}a-d.

In particular, the contribution of the irreducible (connected) correlator of the first and the second terms in
Eq.~(\ref{LSlowFastInteractions}),
\begin{align}
	-\frac{\tilde\varkappa^2}{2}\int\llangle\left[(\psi^\dagger_f\psi_f)(\psi^\dagger_s\psi_s)\right](\br)
	\left[(\psi^\dagger_s\psi_f)(\psi^\dagger_f\psi_s)\right](\br^\prime)\rrangle_f d\br d\br^\prime
	\nonumber\\
	=-\frac{\tilde\varkappa^2}{2}\int (\psi^\dagger_s\psi_s)(\br)
	\psi^\dagger_s(\br^\prime)
	\langle\psi_f(\br^\prime)\psi^\dagger_f(\br)\rangle_f
	\nonumber\\
	\langle\psi_f(\br)\psi^\dagger_f(\br^\prime)\rangle_f
	\psi_s(\br^\prime) d\br d\br^\prime,
	\label{ExampleCorrelator}
\end{align}
equals the diagram in Fig.~\ref{Fig:LineDiagrams}c. Interchanging the expressions
in the square brackets in Eq.~(\ref{ExampleCorrelator})
corresponds then to the diagram \ref{Fig:LineDiagrams}d, which has the same value.
Similarly, the correlator of the second term in Eq.~(\ref{LSlowFastInteractions})
with itself corresponds to the diagram~\ref{Fig:LineDiagrams}b,
of the third and the fourth terms-- to the diagram~\ref{Fig:LineDiagrams}a.
The other correlators vanish.

The four correlators, corresponding to the diagrams Fig.~7a-d, contribute equally to
the renormalisation of the disorder strength $\varkappa$ and lead to the RG flow equation (\ref{gammaflow0}).

{\it Preexponential factor renormalisation.} 
Integrating out the fast fields $\psi_f$ and $\psi_f^\dagger$
renormalises not only the Lagrangian but also the preexponential factor
in the expression for the density of states, Eq.~(\ref{Action1}).

Indeed, %when integrating the fast fields,
due to the correlations between the fast components of the
supersymmetry-breaking preexponential factor $\propto\int\psi_\beta(\br)\psi^\dagger_\beta(\br)d\br
=\int\psi_{s\beta}(\br)\psi^\dagger_{s\beta}(\br)d\br+\int\psi_{f\beta}(\br)\psi^\dagger_{f\beta}(\br)d\br$
and the Lagrangian $\cL_i$, 
the former is renormalised as 
%\equiv\sum_\bk\psi^\dagger_{\bk\beta}\psi_{\bk\beta}$
%is renormalised as
\begin{align}
	\int\psi^\dagger_s(\br)\psi_s(\br)d\br
	\rightarrow
	\int\psi^\dagger_s(\br)\psi_s(\br)d\br
	\nonumber\\
	-\int\left<\psi_{f\beta}(\br)\psi^\dagger_{f\beta}(\br)
	\cL_i(\psi_s^\dagger,\psi_s,\psi_f^\dagger,\psi_f)\right>_fd\br.
	\label{Prefactor}
\end{align}
Using Eqs.~(\ref{LSlowFastInteractions}) we find straightforwardly that the modification
(\ref{Prefactor}) is equivalent to multiplying $\int\psi^\dagger_s(\br)\psi_s(\br)d\br$ by
$1+\delta\lambda/\lambda\equiv\lambda(Ke^{-l})/\lambda(K)$.

Therefore, as a result of integrating out the fast fields the expressions (\ref{Action1})-(\ref{Action3})
reduce to the same form with all the effects of the fast fields encoded in the renormalised parameters
$\lambda$ and $\tilde\varkappa$.

%%%%%%%%%%%%%%%%%%%%%%%%%%%%%BIBLIOGRAPHY%%%%%%%%%%%%%%%%%%%%%%%%%%%

\end{document}